\documentclass[
 reprint,
 aps,
 prr,
 showpacs,
 amsmath,amssymb,
 dvipdfmx,
 longbibliography
]{revtex4-1}

\usepackage{graphicx}
\usepackage{dcolumn}
\usepackage{color}
\usepackage{bm}
\usepackage{here}
\usepackage{subfigure}
\usepackage{siunitx}
\usepackage{tabularx}
\newcolumntype{C}{>{\centering\arraybackslash}X}
\newcolumntype{L}{>{\raggedright\arraybackslash}X}
\newcolumntype{R}{>{\raggedleft\arraybackslash}X}
\usepackage{multirow}

\usepackage[
colorlinks=true, 
pdfstartview=FitV, 
linkcolor=blue, 
citecolor=magenta, 
urlcolor=blue, 
bookmarks=true,
bookmarksnumbered=true,
pdftitle={},
pdfauthor={}
]{hyperref}

\usepackage{amsthm}

\usepackage{comment}
\newcommand{\sectionprl}[1]{{\em #1}\/.---}

\newcommand{\kx}{(k_x)}

\begin{document}
\title{Molecular dynamics study of shear-induced long-range correlations in simple fluids}
\author{Hiroyoshi Nakano$^1$, and Yuki Minami$^2$}
\affiliation{$^1$ Department of Applied Physics and Physico-Informatics, Keio University, Kanagawa 223-8522, Japan}
\affiliation{$^2$Department of Physics, Zhejiang University, Hangzhou 310027, China}

\date{\today}
\begin{abstract}
We investigate long-range correlations (LRCs) induced by shear flow using the molecular dynamics (MD) simulation.
We observe the LRCs by comparing the MD result with the linearized fluctuating hydrodynamics (LFH).
We find that the MD result has large finite-size effects, and it prevents the occurrence of LRCs in small systems.
We examine the finite-size effects using sufficiently large systems consisting of more than ten million particles, and verify the existence of shear-induced LRCs without ambiguity.
Furthermore, we show that MD result is quantitatively consistent with the LFH solution for the large system. 
As we reduce the system size $L$ or increase the shear rate $\dot{\gamma}$, the hydrodynamic description gradually breaks down in the long-wavelength region.
We define a characteristic wavenumber $k^{\rm vio}$ associated with the breakdown and find the nontrivial scaling relations $k^{\rm vio} \propto L^{-\omega}$ and $k^{\rm vio} \propto \dot{\gamma}$, where $\omega$ is an exponent depending on $\dot{\gamma}$. 
These relations enable us to estimate the finite-size effects in a larger-size simulation from a smaller system.
\end{abstract}
\pacs{}
\maketitle

\section{Introduction}
For equilibrium systems with short-range interactions, long-range correlations (LRCs) appear in certain situations, such as for a critical point and for an ordered phase with spontaneous symmetry breaking.  
Near the critical point, the correlation length diverges and the correlation function exhibits a long-range nature~\cite{Goldenfeld2018lectures}.
In the ordered phase with spontaneous breaking of the continuous symmetry, the so-called Nambu--Goldstone mode~\cite{Goldstone1961field,Goldstone1962broken,Nambu1961dynamical} appears and leads to the LRC.

For nonequilibrium systems, the LRCs exist in various situations, even in a disordered phase far from the critical point~\cite{Garrido1990long,Dorfman1994generic,DeZarate2006hydrodynamic}. Extensive theoretical studies since the 1980s have shown that LRCs are a general feature of stationary nonequilibrium systems with conservation laws and anisotropy~\cite{Garrido1990long}. 
In addition, experimental studies have observed LRCs under a temperature gradient~\cite{Law1988light,Segre1992rayleigh,Takacs2008dynamics,Takacs2011thermal}. Recently, nonequilibrium LRCs have attracted attention as the origin of Casimir-like long-range forces~\cite{Kruger2011nonequilibrium,Kirkpatrick2013giant,Aminov2015fluctuation,DeZarate2019nonequilibrium}. Moreover, they have been studied in relation to nonequilibrium phase transitions~\cite{Schmittmann1993fixed,Bassler1995,Tauber2002effects,Bodineau2008long} and in the context of constructing the theoretical framework of nonequilibrium statistical mechanics~\cite{Derrida2002large,Wada2003anomalous,Sasa2006steady,Bertini2007stochastic}.

The mechanism and nature of nonequilibrium LRCs have been established using phenomenological models such as fluctuating hydrodynamics~\cite{Kirkpatrick1982light,Ronis1982nonlinear,Lutsko1985hydrodynamic,Grinstein1990conservation,Wada2004shear} and stochastic lattice gases~\cite{Spohn1983long,Katz1984nonequilibrium,Mahdisoltani2021long}.
These coarse-grained models enable the nonequilibrium fluctuations to be examined in terms of the violation of a detailed-balance condition, conservation law, and anisotropy. However, they do not produce the mechanism of nonequilibrium LRCs from molecular-scale dynamics. 
There are few theoretical attempts to study nonequilibrium LRCs from the underlying Hamiltonian dynamics.
Then, how LRCs arise from the molecuar-scale dynamics remains poorly understood.

In this paper, we study nonequilibrium LRCs in simple fluids under shear flow using molecular dynamics (MD) simulations.
A particle system under shear flow is one of the simplest nonequilibrium setups, and has been used to probe nonequilibrium LRCs in a large number of simulation studies~\cite{Naitoh1978the,Naitoh1979shear,Hoover1980lennard,Evans1989on,Evans1990viscosity,Travis1998strain,Istvan2002shear,Evans1980computer,Erpenbeck1984shear,Marcelli2001analytic,Jlian2001energy,Ahmed2009strain,Varghese2015hydrodynamic,Varghese2017spatial,Jialin2003scaing,Todd2005power,Otsuki2009spatial,Otsuki2009long,Desgranges2009universal,Lautenschlaeger2019shear}. 
However, it is difficult to probe shear-induced LRCs without ambiguity in the MD simulations as reported in the previous studies~\cite{DeZarate2019nonequilibrium}.
We now briefly review the previous studies and the difficulties.

In the hydrodynamic description, nonequilibrium fluctuations consist of two terms:
\begin{eqnarray}
\langle A(\bm{r}) B(\bm{r}') \rangle \sim c_1\delta(\bm{r}-\bm{r}') + \frac{c_2}{|\bm{r}-\bm{r}'|^{\alpha}},
\label{eq:identify the long-range correlations}
\end{eqnarray}
where $A(\bm{r})$ and $B(\bm{r})$ are density fields (e.g., density fluctuations $\delta \rho(\bm{r})$ or velocity fluctuations $\delta v^a(\bm{r})$), and $c_1$ and $c_2$ are appropriate constants. The first term proportional to the delta function implies that $A(\bm{r})$ and $B(\bm{r})$ are uncorrelated on the hydrodynamic scale. Then, the second term represents the LRCs, which are generally absent in equilibrium fluids.

The fluctuating hydrodynamics provides a phenomenological model for describing the fluctuations at the hydrodynamic scale. This model is widely used to study shear-induced LRCs. 
One characteristic behavior of shear-induced LRCs is the crossover between two power-law decays~\cite{Lutsko1985hydrodynamic,Wada2003anomalous}. For example, the spatial correlation of the density fluctuation $\langle \delta \rho(\bm{r}) \delta \rho(\bm{r}') \rangle$ decays according to $|\bm{r}-\bm{r}'|^{-1}$ for short-distance scales and crosses over to the stronger decay $|\bm{r}-\bm{r}'|^{-11/3}$ for long-distance scales. Similarly, the spatial correlation of the velocity fluctuations $\langle \bm{v}(\bm{r})\cdot \bm{v}(\bm{r}') \rangle$ crosses over from $|\bm{r}-\bm{r}'|^{-1}$ to $|\bm{r}-\bm{r}'|^{-5/3}$.

Another important prediction from the fluctuating hydrodynamics is the existence of shear-induced corrections to the pressure $P$ and shear viscosity $\eta$. These corrections arise from the nonlinear coupling of the LRCs. 
Kawasaki and Gunton initially found these corrections by using the projection operator method and the mode-coupling theory~\cite{Kawasaki1973theory}.  
They were subsequently derived from the fluctuating hydrodynamics~\cite{Lutsko1985mode, Wada2003anomalous,DeZarate2019nonequilibrium}.
The shear-induced correction to the pressure $P$ depends on the Reynolds number $Re$, and is given in the two limits as
\begin{eqnarray}
\begin{array}{c}
P - P_{\rm eq} \propto L\dot{\gamma}^{2} \ \ {\rm for} \ Re \ll 1, \\[3pt]
P - P_{\rm eq} \propto  \dot{\gamma}^{3/2} \ \ {\rm for} \ Re \gg 1,
\end{array}
\label{eq:non-analytical shear-rate dependence of pressure}
\end{eqnarray}
where $P$, $L$, and $\dot{\gamma}$ are the pressure, system size, and shear rate, and $P_{\rm eq}$ is the pressure in the limit $\dot{\gamma}\to 0$. 
For the shear viscosity, the corresponding behavior is given by
\begin{eqnarray}
\eta - \eta_{\rm eq} \propto \dot{\gamma}^{1/2},
\label{eq:non-analytical shear-rate dependence of viscosity}
\end{eqnarray}
where $\eta$ is the viscosity and $\eta_{\rm eq}$ is the viscosity in the limit $\dot{\gamma}\to 0$.

After these results had been obtained by the fluctuating hydrodynamics or kinetic theory, numerous MD simulations attempted to verify them. 
The basic idea was to probe the shear-induced LRCs by observing Eq.~(\ref{eq:non-analytical shear-rate dependence of pressure}) or (\ref{eq:non-analytical shear-rate dependence of viscosity}). The results remain controversial. Earlier simulation results~\cite{Naitoh1978the,Naitoh1979shear,Hoover1980lennard} were interpreted in favor of the non-analytical shear-rate dependence. In particular, Evans and coworkers~\cite{Evans1989on,Evans1990viscosity,Travis1998strain,Istvan2002shear} calculated the shear viscosity at the Lennard--Jones triple point and observed Eq.~(\ref{eq:non-analytical shear-rate dependence of viscosity}). However, more sophisticated simulations~\cite{Marcelli2001analytic,Jlian2001energy,Ahmed2009strain,Varghese2015hydrodynamic,Varghese2017spatial} support the assertion that the shear-induced correction behaves as $\dot{\gamma}^2$, not as $\dot{\gamma}^{3/2}$. Furthermore, the MD simulations of Sadus and coworkers~\cite{Jialin2003scaing,Todd2005power} found that the exponent of pressure varies continuously between $1.2$ and $2.0$ depending on the temperature and density.
More recently, Ortiz de Z\'{a}rate et al.~\cite{DeZarate2019nonequilibrium} reported that the shear-induced correction has two different origins, from short- and long-range scales. The long-range correction comes from the nonlinear coupling of the LRCs, which is calculated by the fluctuating hydrodynamics. The short-range correction is a molecular-scale effect and is independent of the LRCs. Ortiz de Z\'{a}rate et al.~estimated the magnitude of the short-range correction using kinetic theory and demonstrated that it yields non-negligible contributions. Their argument suggests the possibility that previous MD simulations captured the short-range correction. Thus, we find it difficult to extract the shear-induced LRCs from the shear-rate dependence of pressure $P$ and shear viscosity $\eta$.

Another direction for probing the existence of LRCs is through direct observations, such as Eq.~(\ref{eq:identify the long-range correlations}). 
Two groups studied the LRCs along this direction: Otsuki and Hayakawa~\cite{Otsuki2009long,Otsuki2009spatial} and Varghese et al.~\cite{Varghese2015hydrodynamic,Varghese2017spatial}. 
Otsuki and Hayakawa initially succeeded in observing the power-law decay of density and velocity fluctuations in a granular particle system, and found that the exponent $\alpha$ is close to the value predicted by the fluctuating hydrodynamics~\cite{Otsuki2009spatial}. Their simulation size was insufficient for quantitatively examination of large-distance correlations beyond $10\sigma$, where $\sigma$ is the diameter of the particles. Subsequently, Varghese et al.~performed a mesoscale simulation based on the multiparticle collision dynamics~\cite{Varghese2017spatial}. They successfully observed the shear-induced LRCs, and reported the behavior that is quantitatively consistent with the fluctuating hydrodynamics. However, the multiparticle collision dynamics is not based on microscopic interactions and cannot describe the molecular-scale behavior.

In this paper, we directly observe the LRCs by comparing the MD results with the linearized fluctuating hydrodynamics (LFH).
We find that the MD result has large finite-size effects, and it prevents the occurrence of LRCs in small systems.
We examine the finite-size effect using a sufficiently large system consisting of more than ten million particles, and show the existence of shear-induced LRCs without ambiguity.

Furthermore, we verify that our MD result is quantitatively consistent with the LFH solution for the large system. 
However, as we reduce the system size or increase the shear rate, the MD result gradually deviates from the LFH solution in the long-wavelength region. 
As a quantitative description of how the deviation increases, we define the characteristic wavenumber $k^{\rm vio}$ such that the prediction from the fluctuating hydrodynamics is valid for $k>k^{\rm vio}$. We find that $k^{\rm vio}$ has a nontrivial scaling dependence on the system size and shear rate.

The remainder of this paper is organized as follows. In Sec.~\ref{sec:Phenomenological description of shear-induced long-range correlations}, we briefly review the analysis results based on the fluctuating hydrodynamics. In Sec.~\ref{sec:Setup of molecular dynamics simulation}, we explain the setup of the MD simulations. The main part of this paper is Sec.~\ref{sec:Main results}, where the MD result is presented and compared with the LFH solution. Section~\ref{sec:Discussion} gives our concluding remarks and discussions.

\section{Hydrodynamic description of shear-induced long-range correlations}
\label{sec:Phenomenological description of shear-induced long-range correlations}
The fluctuating hydrodynamics provides a powerful analytical tool for describing the nonequilibrium LRCs. Here, we briefly review the established results regarding shear-induced LRCs.

\subsection{Model}
\label{subsec: hydrodynamic model}
We consider an isothermal fluid with a uniform temperature $T$ defined in a three-dimensional region $[-L_x/2,L_x/2]\times [-L_y/2,L_y/2] \times [-L_z/2,L_z/2]$.
The isothermal fluid is described by two fluctuating fields, namely the density $\rho(\bm{r},t)$ and the velocity $\bm{v}(\bm{r},t)$. The time evolution of $\rho(\bm{r},t)$ and $\bm{v}(\bm{r},t)$ is given by~\cite{landau1959fluid}
\begin{eqnarray}
\frac{\partial \rho}{\partial t} + \frac{\partial}{\partial x_l} (\rho v_l)= 0,
\label{eq: basic EOM0}\\
\frac{\partial}{\partial t} (\rho v_i) + \frac{\partial \Pi_{ij}}{\partial x_j} = 0,
\label{eq: basic EOM1}
\end{eqnarray}
where $\Pi_{ij}(\bm{r},t)$ is the momentum flux tensor, written as 
\begin{eqnarray}
\Pi_{ij} &=& \rho v_i v_j + p \delta _{ij} - \eta_0 \biggl(\frac{\partial v_j}{\partial x_i} + \frac{\partial v_i}{\partial x_j} - \frac{2}{3} \delta_{ij}\frac{\partial v_l}{\partial x_l} \biggr) \nonumber \\
&-& \zeta_0 \delta_{ij} \frac{\partial v_l}{\partial x_l} + s_{ij}.
\end{eqnarray}
Here, $\eta_0$ is the bare shear viscosity, $\zeta_0$ is the bare bulk viscosity, $p(\bm{r},t)$ is the pressure, and $s_{ij}(\bm{r},t)$ is the Gaussian random noise tensor satisfying
\begin{eqnarray}
\langle s_{ik}(\bm{r},t) s_{lm}(\bm{r}',t') \rangle = 2 T \eta_{iklm} \delta^3(\bm{r}-\bm{r}') \delta(t-t'),
\label{eq: basic EOM2} \\
\eta_{iklm} = \eta_0 \delta_{il} \delta_{km} + \eta_0\delta_{im} \delta_{kl} + \biggl(\zeta_0-\frac{2}{3}\eta_0\biggr)\delta_{ik} \delta_{lm}.
\end{eqnarray}

\begin{figure}[t]
\centering
\begin{center}
\includegraphics[width=6.4cm]{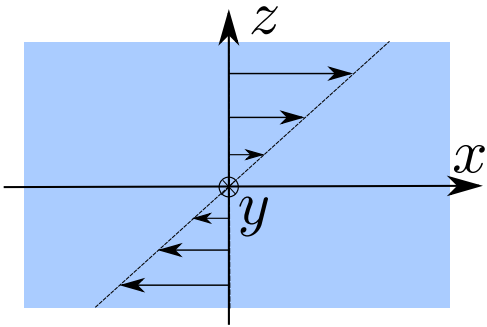} 
\end{center}
\caption{Schematic illustration of shear flow. The $x$-axis is in the direction of the flow velocity and the $z$-axis is in the direction of the velocity gradient. }
\label{fig: illustration of shear flow}
\end{figure}
We study the nonequilibrium steady state characterized by the average density field and velocity field. i.e.,
\begin{eqnarray}
\langle \rho(\bm{r}) \rangle = \rho_0, \; \langle \bm{v}(\bm{r}) \rangle = (\dot{\gamma} z ,0,0).
\end{eqnarray}
A schematic illustration of the steady state is presented in Fig.~\ref{fig: illustration of shear flow}.
In analyzing the fluctuating hydrodynamics, we focus on the bulk region and neglect boundary effects.  
Therefore, we do not have to specify the boundary condition. 
This is crucially different from the setup adopted in the MD simulations. Boundary effects are inevitable in the MD simulations because the nonequilibrium steady state is maintained using the Lees--Edwards boundary condition as explained in the next section. 
We will revisit this difference in Sec.~\ref{sec:Main results}, where we compare the MD result with the LFH solution.

\subsection{Spatial correlation of momentum field}
\label{subsec:spatial correlation of momentum field}
In the fluctuating hydrodynamics, the spatial correlation of the momentum is defined as
\begin{eqnarray}
C^{\rm FH}_{ij}(\bm{r},\bm{r}') = \langle \delta g_i(\bm{r},t) \delta g_j(\bm{r}',t) \rangle,
\end{eqnarray}
where $\delta g_i(\bm{r},t)$ is given by
\begin{eqnarray}
\delta \bm{g}(\bm{r},t) = \rho(\bm{r},t) (\bm{v}\big(\bm{r},t) -\langle\bm{v}(\bm{r},t) \rangle \big).
\label{eq: fluctuation of momentum density field: definition}
\end{eqnarray}
In the MD simulations, we define a counterpart of this correlation function in terms of phase-space variables. 
To avoid confusion, we introduce the superscript $\rm FH$ to denote the fluctuating hydrodynamics.
The existence of nonequilibrium LRCs is identified by the power-law decay of the correlation function. 

The steady state under shear flow has translational symmetry~\cite{Onuki1979nonequilibrium}. This is expressed in terms of the correlation function as $C^{\rm FH}_{ij}(\bm{r},\bm{r}') = C^{\rm FH}_{ij}(\bm{r}+\bm{a},\bm{r}'+\bm{a})$, where $\bm{a}$ is an arbitrary constant vector. 
It is useful to introduce the Fourier transform of the correlation function 
\begin{eqnarray}
C^{\rm FH}_{ij}(\bm{r},\bm{r}') = \int \frac{d^3\bm{k}}{(2\pi)^3} C^{\rm FH}_{ij}(\bm{k}) e^{-i\bm{k}\cdot (\bm{r}-\bm{r}')}.
\end{eqnarray}
We restrict our interest to the two correlation functions $C^{\rm FH}_{xx}(\bm{k})$ and $C^{\rm FH}_{zz}(\bm{k})$ at $k_y=k_z=0$, and denote these as $C^{\rm FH}_{xx}\kx$ and $C^{\rm FH}_{zz}\kx$. 
Note that linear approximations of $C^{\rm FH}_{yy}(k_x)$ are not affected by the shear flow~\cite{Lutsko1985hydrodynamic,Varghese2017spatial}. Therefore, we do not discuss $C^{\rm FH}_{yy}(k_x)$ in this paper.

From Eqs.~(\ref{eq: basic EOM0}) and (\ref{eq: basic EOM1}), we derive the integral expressions for $C^{\rm FH}_{xx}\kx$ and $C^{\rm FH}_{zz}\kx$ under the linear approximations:
\begin{eqnarray}
C^{\rm FH}_{xx}\kx &=& T \rho_0 \nonumber \\[3pt]
& &  \hspace{-1.2cm} -\dot{\gamma}^2 T \rho_0 \int_0^{\infty}ds \frac{s}{(1+\dot{\gamma}^2 s^2)^{3/2}} e^{-\Gamma_0 k_x^2(s+\frac{1}{3} \dot{\gamma}^2 s^3)}, 
\label{eq:correlation function: integral expression: xx} \\
C^{\rm FH}_{zz}\kx &=& T \rho_0  \nonumber \\[3pt]
& &  + 2 \dot{\gamma}^2 T \rho_0 \int_0^{\infty}ds s e^{-2\nu_0 k_x^2(s+\frac{1}{3} \dot{\gamma}^2 s^3)},
\label{eq:correlation function: integral expression: zz}
\end{eqnarray}
with
\begin{eqnarray}
\nu_0 = \frac{\eta_0}{\rho_0}  \ , \ \Gamma_0 = \frac{\zeta_0+4\eta_0/3}{\rho_0}.
\end{eqnarray}
We call Eqs.~(\ref{eq:correlation function: integral expression: xx}) and (\ref{eq:correlation function: integral expression: zz}) the LFH solution. 
$C^{\rm FH}_{xx}\kx$ and $C^{\rm FH}_{zz}\kx$ correspond to the longitudinal and transverse momentum fluctuations, respectively. 
Therefore, Eq.~(\ref{eq:correlation function: integral expression: zz}) for $C^{\rm FH}_{zz}\kx$ does not contain $\zeta_0$, and is the same as that for an incompressible fluid. In contrast, $C^{\rm FH}_{xx}\kx$ is strongly affected by the compressibility of the fluid.
These expressions were initially derived in Ref.~\cite{Lutsko1985hydrodynamic}. Appendix~\ref{appendix: Brief sketch of derivation} provides a brief sketch of the derivation; for further details, see Ref.~\cite{Otsuki2009spatial}.

From the LFH solution of Eqs.~(\ref{eq:correlation function: integral expression: xx}) and (\ref{eq:correlation function: integral expression: zz}), we can see the existence of the shear-induced LRCs. 
First, as $\dot{\gamma} \to 0$, Eqs.~(\ref{eq:correlation function: integral expression: xx}) and (\ref{eq:correlation function: integral expression: zz}) reduce to 
\begin{eqnarray}
C^{\rm FH}_{xx}\kx=C^{\rm FH}_{zz}\kx = T\rho_0.  \label{eq:equilibrium value}
\end{eqnarray}
This means that the correlation in the real space is given by the delta function. The correlation length is interpreted to be of the molecular scale.
For $\dot{\gamma} >0$,  Eqs.~(\ref{eq:correlation function: integral expression: xx}) and (\ref{eq:correlation function: integral expression: zz}) have nonequilibrium corrections, which lead to the LRCs. 
The asymptotic expression of Eq.~(\ref{eq:correlation function: integral expression: zz}) in the long-wavelength region is calculated as 
\begin{eqnarray}
C^{\rm FH}_{zz}\kx = T \rho_0 \Big(1+\frac{1}{2}\frac{\dot{\gamma}^2}{\nu_0^2 k_x^4} \Big)
\label{eq:asymptotic expression for Czz1}
\end{eqnarray}
for $k_x \gg k_x^{\rm cross}$, and 
\begin{eqnarray}
\hspace{-0.5cm} C^{\rm FH}_{zz}\kx = T \rho_0 \Biggl(1+\biggl(\frac{2}{3}\biggr)^{1/3}\Gamma\biggl(\frac{2}{3}\biggr)\frac{\dot{\gamma}^{2/3}}{\nu_0^{4/3} k_x^{4/3} }\Biggr)
\label{eq:asymptotic expression for Czz2}
\end{eqnarray}
for $k_x \ll k_x^{\rm cross}$.
Here, $k_x^{\rm cross}$ determines the crossover scale between Eqs.~(\ref{eq:asymptotic expression for Czz1}) and (\ref{eq:asymptotic expression for Czz2}), and is given by
\begin{eqnarray}
k_x^{\rm cross} = \Biggl(\frac{3}{16}\Gamma\biggl(\frac{2}{3}\biggr)^3\Biggr)^{1/8}\sqrt{\frac{\dot{\gamma}}{\nu_0}}.
\end{eqnarray}
These expressions imply that an additional correlation proportional to $k_x^{-4}$ appears at short-distance scales and crosses over to $k_x^{-4/3}$ at large-distance scales. 
Such power-law behavior in the Fourier space corresponds to an algebraic decay in the real space. 
We can repeat the same discussion for $C^{\rm FH}_{xx}\kx$ and derive the LRC~\cite{Lutsko1985hydrodynamic}.

We use the LFH solution of Eqs.~(\ref{eq:correlation function: integral expression: xx}) and (\ref{eq:correlation function: integral expression: zz}) to probe the existence of shear-induced LRCs in the MD simulation. Additionally, we quantitatively examine the validity of the LFH solution.
Note that expressions such as Eqs.~(\ref{eq:correlation function: integral expression: xx}) and (\ref{eq:correlation function: integral expression: zz}) provide a starting point for explaining various phenomena coming from shear-induced LRCs.
For example, Lutsko and Dufty~\cite{Lutsko1985mode} derived a nonequilibrium correction to the shear viscosity in the form of Eq.~(\ref{eq:non-analytical shear-rate dependence of viscosity}). Similarly, Wada and Sasa~\cite{Wada2003anomalous} and Ortiz de Z\'{a}rate et al.~\cite{DeZarate2019nonequilibrium} derived the shear-rate dependence of pressure for incompressible fluids.
Therefore, it is important to establish the LFH solution quantitatively from the molecular-scale dynamics.

\section{Setup of MD simulations}
\label{sec:Setup of molecular dynamics simulation}
\subsection{Model}
We consider an $N$ particle system that is confined in a three-dimensional region $[-L_x/2,L_x/2]\times [-L_y/2,L_y/2] \times [-L_z/2,L_z/2]$. The dynamics is given by
\begin{eqnarray}
\frac{d\bm{r}_i}{dt} &=& \frac{\bm{p}_i}{m}, \\[3pt]
\frac{d\bm{p}_i}{dt} &=& - \frac{\partial U}{\partial \bm{r}_i} + \bm{f}^{\rm th}_i,
\end{eqnarray}
where $(\bm{r}_i,\bm{p}_i)$ is the position and momentum of the $i$th particle, $m$ is the mass, $U(r)$ is the interparticle interaction, and $\bm{f}^{\rm th}_i$ is the force acting on the $i$th particle from a thermostat. We use the Weeks--Chandler--Andersen (WCA) potential as the interparticle interaction, which is the Lennard--Jones potential with the cutoff-length $r^{\rm LJ}_c=2^{1/6}\sigma$, i.e.,
\begin{eqnarray}
U(r) = 4\epsilon \Big\{\Big(\frac{\sigma}{r} \Big)^{12} - \Big(\frac{\sigma}{r} \Big)^{6} +\frac{1}{4} \Big \} \theta(2^{1/6}\sigma-r),
\label{eq:Lennard-Jones potential}
\end{eqnarray}
where $\theta(r)$ is the Heaviside step function and $\sigma$ is the diameter of the particle.

To maintain a constant temperature under the shear flow, we use the dissipative particle dynamics (DPD) thermostat~\cite{Espa1995statistical}, which is given by
\begin{eqnarray}
\bm{f}^{\rm th}_i = \sum_{j\neq i} \Big[&-&\gamma \omega^D(r_{ij}) (\hat{\bm{r}}_{ij}\cdot \bm{v}_{ij} )\hat{\bm{r}}_{ij} \nonumber \\
&+& \sqrt{2 \gamma k_B T \omega^D(r_{ij})} \theta_{ij} (t)\hat{\bm{r}}_{ij}\Big].
\end{eqnarray}
Here, $\hat{\bm{r}}_{ij}$ is a unit vector in the direction $\bm{r}_{ij} = \bm{r}_i-\bm{r}_j$, $\omega^D(r_{ij})$ is the cutoff function
\begin{eqnarray}
\begin{cases}
\omega^D(r) = 1 - r/r^{\rm DPD}_c& \ {\rm for} \ r<r^{\rm DPD}_c, \\
\omega^D(r) = 0 & \ {\rm for} \ r\geq r^{\rm DPD}_c,
\end{cases}
\end{eqnarray}
and $\theta_{ij}(t)$ is random noise satisfying $\langle \theta_{ij}(t) \theta_{kl}(t') \rangle = (\delta_{ik}\delta_{jl}+\delta_{il}\delta_{jk})\delta(t-t')$. $\gamma$ and $T$ represent the friction and the temperature of thermostat, respectively. Because the DPD thermostat satisfies the fluctuation-dissipation relation, our model relaxes to equilibrium when no external forces are imposed. The cutoff length $r^{\rm DPD}_c$ is set to $2.0\sigma$. 

Note that the DPD thermostat obeys Newton's third law, which ensures momentum conservation~\cite{Groot1997dissipative}. This is why we apply the DPD thermostat. Indeed, one of the origins of nonequilibrium LRCs is the conservation law~\cite{Garrido1990long,Dorfman1994generic}.

The shear flow is realized using the Lees--Edwards boundary condition~\cite{Lees1972the,allen2017computer} along the $z$-axis. Along the $x$- and $y$-axes, we impose standard periodic boundary conditions. Thus, the velocity profile in the steady state is realized as
\begin{eqnarray}
\tilde{\bm{v}}(\bm{r}) = (\dot{\gamma} z, 0,0).
\label{eq:velocity profile in the steady state}
\end{eqnarray}
Note that the Lees--Edwards boundary condition violates the momentum conservation law along the $x$-direction. The total amount of momentum along the $x$-direction depends on the number of atoms that leave the lower and upper boundaries. 
This is because, when one atom leaves the lower (upper) boundary $z=-L_z/2$ ($z=L_z/2$) with velocity $\bm{u}$, the corresponding atom is introduced from the upper (lower) boundary with velocity $\bm{u}\pm \dot{\gamma}L_z\hat{\bm{e}}_x$.
However, except at the boundaries, the local conservation law still holds. Moreover, in the steady state, because the net mass transfer via the lower or upper boundary is balanced, the violation is sufficiently small and the time-averaged total momentum must be zero. Thus, we expect that the effect of the violating the conservation law through the Lees--Edwards boundary condition will be sufficiently small~\footnote{
Another boundary condition that allows the bulk fluid to produce uniform shear flow is the movement of two solid walls parallel to each other. This boundary condition also violates the conservation law of total momentum parallel to the walls~\cite{Bocquet1994hydrodynamic}
}.

\subsection{Parameters}
In the numerical simulations, all quantities are measured by the Lennard--Jones units $(m,\sigma,\epsilon)$. In particular, the time is measured by $\tau_{\rm unit} = \sqrt{m\sigma^2/\epsilon}$. All the MD simulations are performed by LAMMPS (Large-scale Atomic/Molecular Massively Parallel Simulator)~\cite{Plimpton1995fast,Thompson2022LAMMPS}. The time integration is calculated by the velocity Verlet algorithm. The timestep is set to $0.0025$, $0.00375$, or $0.005$ depending on the shear rate and the system size. We fix the temperature and friction of the thermostat to $T=1.0$ and $\gamma=1.0$, respectively. The density $\rho_0 \equiv N/L_xL_yL_z$ is fixed to $0.78$.

The transport coefficients take almost the same value in all simulations. In particular, we use $\eta_0 = 1.74$ and $\zeta_0=14.04$ to compare the MD result with the LFH solution, which are calculated from the Green--Kubo formula (see Appendix~\ref{appendix:Measurement of viscosity} for details).

\subsection{Observation method}
All observations are performed in the nonequilibrium steady state, which is prepared by different methods depending on the system size $L_z$. For $L_z \leq 512$, we start from the initial state in which the particles are randomly located with zero overlaps.
We then perform the relaxation run for about 10 times the relaxation time. The relaxation time is estimated from the relaxation of the velocity profile (see Appendix~\ref{appendix:Measurement of relaxation time} for details). For $L_x=1024$, $L_y=32$, and $L_z=512$ ($N=\num{13086228}$), the relaxation time is about $3000$, and the relaxation run with a timestep $0.0025$ takes $100$ hours using 16 nodes of the ISSP supercomputer (AMD EPYC 7702, 64 cores $\times$ 2 per node). 

For $L_z > 512$, we adopt a locally relaxed state as the initial state and perform the relaxation run for about three times the relaxation time. For example, for $L_x=1024$, $L_y=32$, and $L_z=1024$, the initial state is prepared by combining two different relaxation states for $L_x=1024\sigma$, $L_y=32\sigma$, and $L_z=512\sigma$.

After the relaxation run, we observe the correlation function of the momentum fluctuation:
\begin{eqnarray}
C^{\rm MD}_{ij}(\bm{k};\dot{\gamma}) = \frac{1}{V} \langle\delta \tilde{g}^i(\bm{k})\delta \tilde{g}^j(-\bm{k}) \rangle,
\end{eqnarray}
where $\langle \cdot \rangle$ represents the time average in the steady state and the ensemble average over different noise realizations.
Here, $\delta\hat{g}^i(\bm{k})$ ($i=x,y,z$) is the Fourier transform of the momentum density field with the mean flow subtracted, which is expressed as
\begin{eqnarray}
\delta \tilde{\bm{g}}(\bm{k}) &=& \int d^3\bm{r} \delta \tilde{\bm{g}}(\bm{r}) e^{-i\bm{k}\cdot \bm{r}}\nonumber \\[3pt]
&=&\sum_{i=1}^N [\bm{p}_i-m\dot{\gamma} z_i \hat{\bm{e}}_x] e^{-i\bm{k}\cdot \bm{r}_i}
\label{eq:def of microscopic momentum fluctuaiton: fourier}
\end{eqnarray}
with
\begin{eqnarray}
\delta \tilde{\bm{g}}(\bm{r}) = \sum_{i=1}^N [\bm{p}_i-m\dot{\gamma} z_i \hat{\bm{e}}_x] \delta(\bm{r}-\bm{r}_i).
\label{eq:def of microscopic momentum fluctuaiton: real}
\end{eqnarray}
We rewrite $\delta \tilde{\bm{g}}(\bm{r})$ in terms of the microscopic density field $\tilde{\rho}(\bm{r}) = \sum_{i=1}^N m\delta(\bm{r}-\bm{r}_i)$ and momentum field $\tilde{\bm{g}}(\bm{r}) = \sum_{i=1}^N \bm{p}_i\delta(\bm{r}-\bm{r}_i)$ as
\begin{eqnarray}
\delta \tilde{\bm{g}}(\bm{r}) &=& \tilde{\bm{g}}(\bm{r}) - \tilde{\rho}(\bm{r})\langle \tilde{\bm{v}}(\bm{r}) \rangle.
\label{eq:rewrite of microscopic momentum fluctuaiton}
\end{eqnarray}
By comparing Eq.~(\ref{eq:rewrite of microscopic momentum fluctuaiton}) to Eq.~(\ref{eq: fluctuation of momentum density field: definition}), we find that $C^{\rm MD}_{ij}(\bm{k};\dot{\gamma})$ is the microscopic counterpart of $C^{\rm FH}_{ij}(\bm{k};\dot{\gamma})$.

We also introduce the relative deviation $\Delta_{ij}\kx$ to verify the validity of the LFH solution: 
\begin{eqnarray}
\Delta_{ij}\kx = \Biggl|\frac{C^{\rm MD}_{ij}\kx-C^{\rm FH}_{ij}\kx}{C^{\rm MD}_{ij}\kx}\Biggr|.
\end{eqnarray}
In a region where the relative deviation is large, the LFH solution cannot be applied to describe the MD result. 
For a quantitative discussion, we introduce the criterion $\Delta_{ij}\kx < 0.1$ for the applicability of the LFH solution.
We then define the characteristic wavenumber $k_x^{\rm vio}$ as the largest wave number satisfying $\Delta_{ij}\kx>0.1$. For the wavenumber region $k_x > k_x^{\rm vio}$, the descriptions given by the fluctuating hydrodynamics are quantitatively valid.

\section{Main results}
\label{sec:Main results}
\subsection{Nonequilibrium LRC}
\label{subsec:Nonequilibrium long-range correlation}

\begin{figure*}[t]
\centering
\begin{tabular}{cc}
\begin{minipage}{0.33\hsize}
\begin{center}
\includegraphics[width=5.8cm]{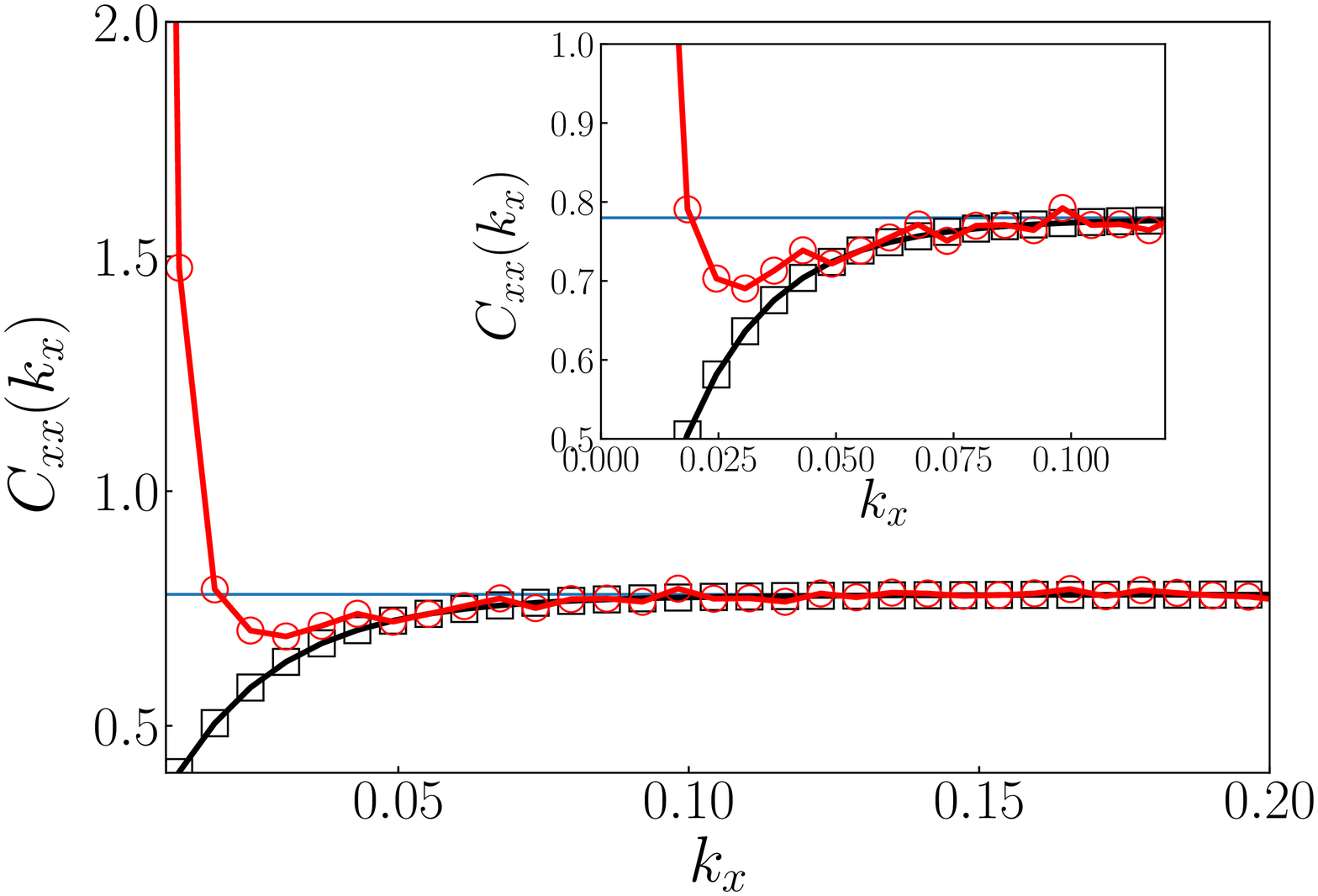} 
\end{center}
\end{minipage}
\begin{minipage}{0.33\hsize}
\begin{center}
\includegraphics[width=5.8cm]{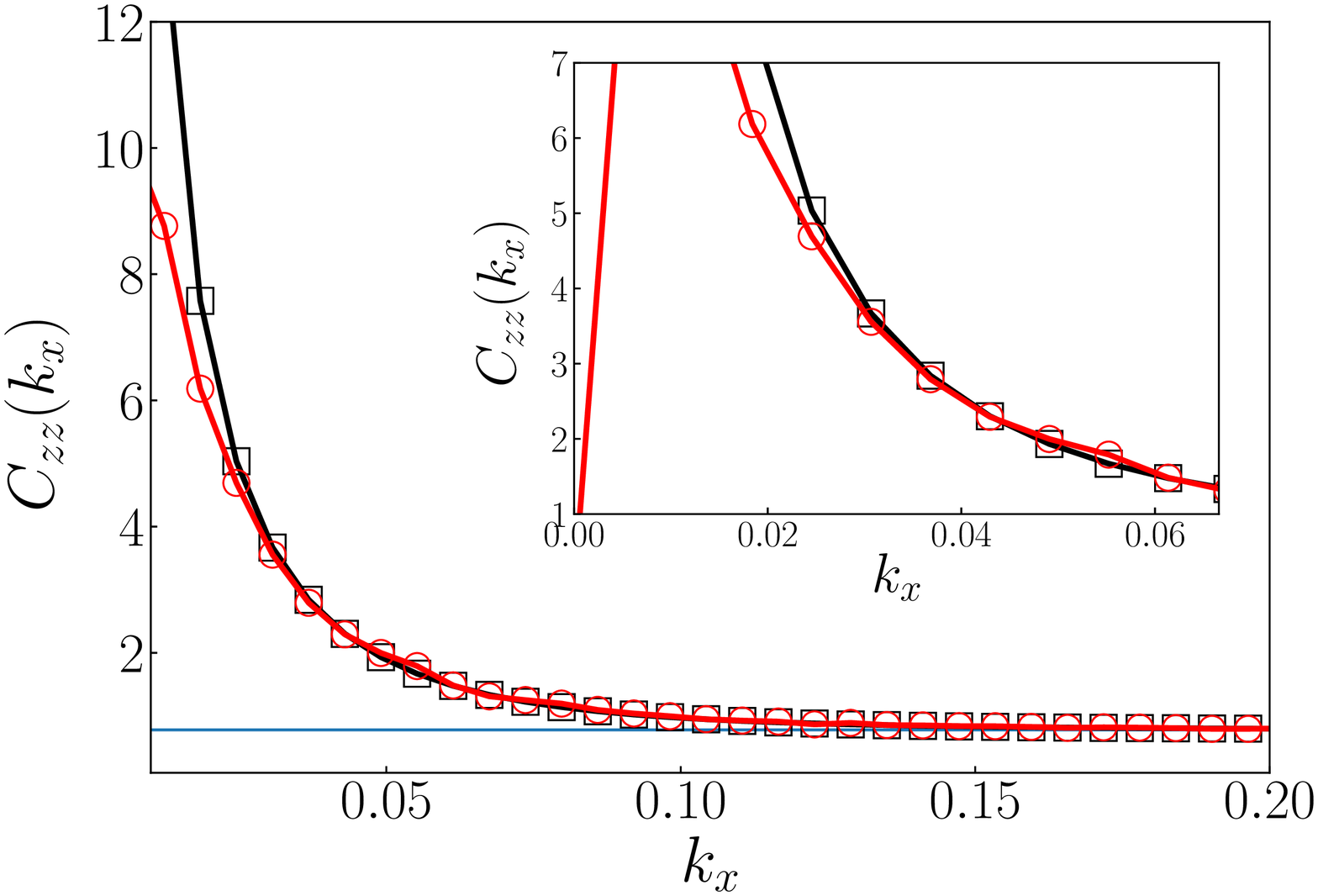} 
\end{center}
\end{minipage}
\begin{minipage}{0.33\hsize}
\begin{center}
\includegraphics[width=5.8cm]{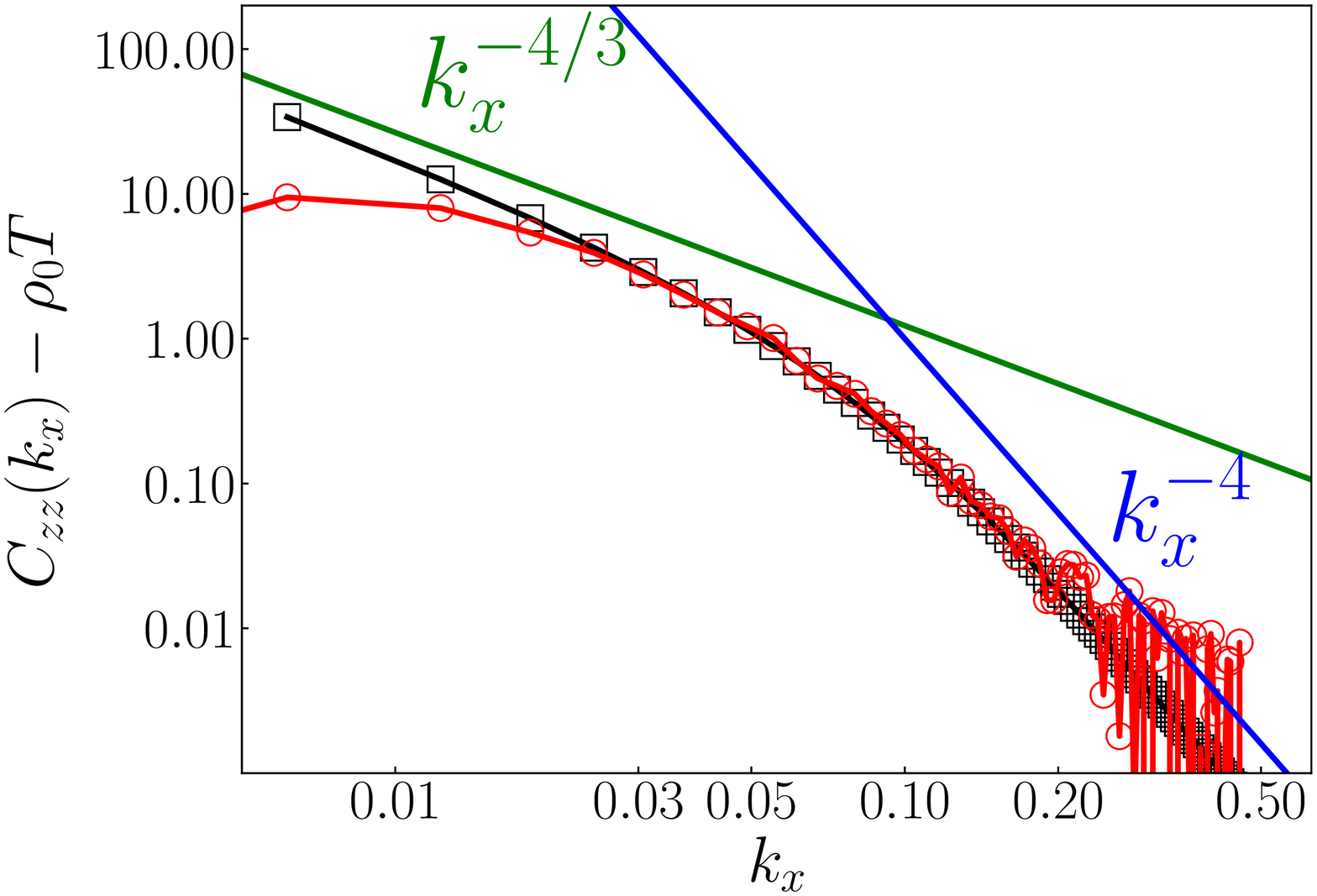} 
\end{center}
\end{minipage}
\end{tabular}
\caption{Left: linear plot of $C_{xx}\kx$. Middle: linear plot of $C_{zz}\kx$. Right: double-log plot of $C_{zz}\kx-\rho_0 T$. The parameters are set to $\dot{\gamma}=0.02$, $L_x=1024$, $L_y=32$, and $L_z=1024$ ($N=\num{26172456}$). The red line is the MD result and the black line is the LFH solution. In the left and middle panels, the blue lines give the equilibrium value from Eq.~(\ref{eq:equilibrium value}). The insets show enlargements of the long-wavelength region. In the right panel, the green and blue lines indicate the two power-law dependences $k_x^{-4/3}$ and $k_x^{-4}$.}
\label{fig: comparison between MD and LFH: largest system}
\end{figure*}

Figure~\ref{fig: comparison between MD and LFH: largest system} presents the results for $\dot{\gamma}=0.02$, $L_x=1024$, $L_y=512$, and $L_z=1024$ ($N=\num{26172456}$), which is the largest system size that we examined. 
The blue lines in the left- and middle-hand panels represent the equilibrium value from Eq.~(\ref{eq:equilibrium value}), and the deviations from this value give the shear-induced correction.  
The black lines show the LFH solutions of Eqs.~(\ref{eq:correlation function: integral expression: xx}) and (\ref{eq:correlation function: integral expression: zz}).
The MD result clearly exhibits shear-induced corrections, and is in quantitative agreement with the LFH solution except in the long-wavelength region. 

We can identify the nonequilibrium LRC from the power-law behavior of Eqs.~(\ref{eq:asymptotic expression for Czz1}) and (\ref{eq:asymptotic expression for Czz2}), as explained in Sec.~\ref{subsec:spatial correlation of momentum field}.
The right-hand panel of Fig.~\ref{fig: comparison between MD and LFH: largest system} shows double-log plots of $C_{zz}^{\rm MD}\kx$ and $C_{zz}^{\rm FH}\kx$ (red and black lines, respectively).
From the LFH solution, the crossover scale between the $k_x^{-4}$ and $k_x^{-4/3}$ behavior is $k_x^{\rm cross} = 0.086$. The MD result is quantitatively consistent with the LFH solution around the crossover region. Thus, we conclude that the MD result exhibits the nonequilibrium LRC. 

In the long-wavelength region, there is a qualitative difference between $C_{xx}^{\rm MD}\kx$ and $C_{xx}^{\rm FH}\kx$.
Specifically, as shown in the left-hand panel of Fig.~\ref{fig: comparison between MD and LFH: largest system}, the MD result monotonically increases from the equilibrium value as $k_x \to 0$, whereas the LFH solution monotonically decreases from the equilibrium value. 
We expect that the boundary effect has a strong influence on the long-wavelength behavior, and thus causes this difference. 
The boundary effect is neglected in the LFH solution, as explained in Sec.~\ref{subsec: hydrodynamic model}.
We now study how the MD result is affected by changes in the shear rate and system size.

\subsection{System-size dependence of nonequilibrium LRC}
\begin{figure}[tb]
\centering
\begin{tabular}{cc}
\begin{minipage}{0.5\hsize}
\begin{center}
\includegraphics[width=4.3cm]{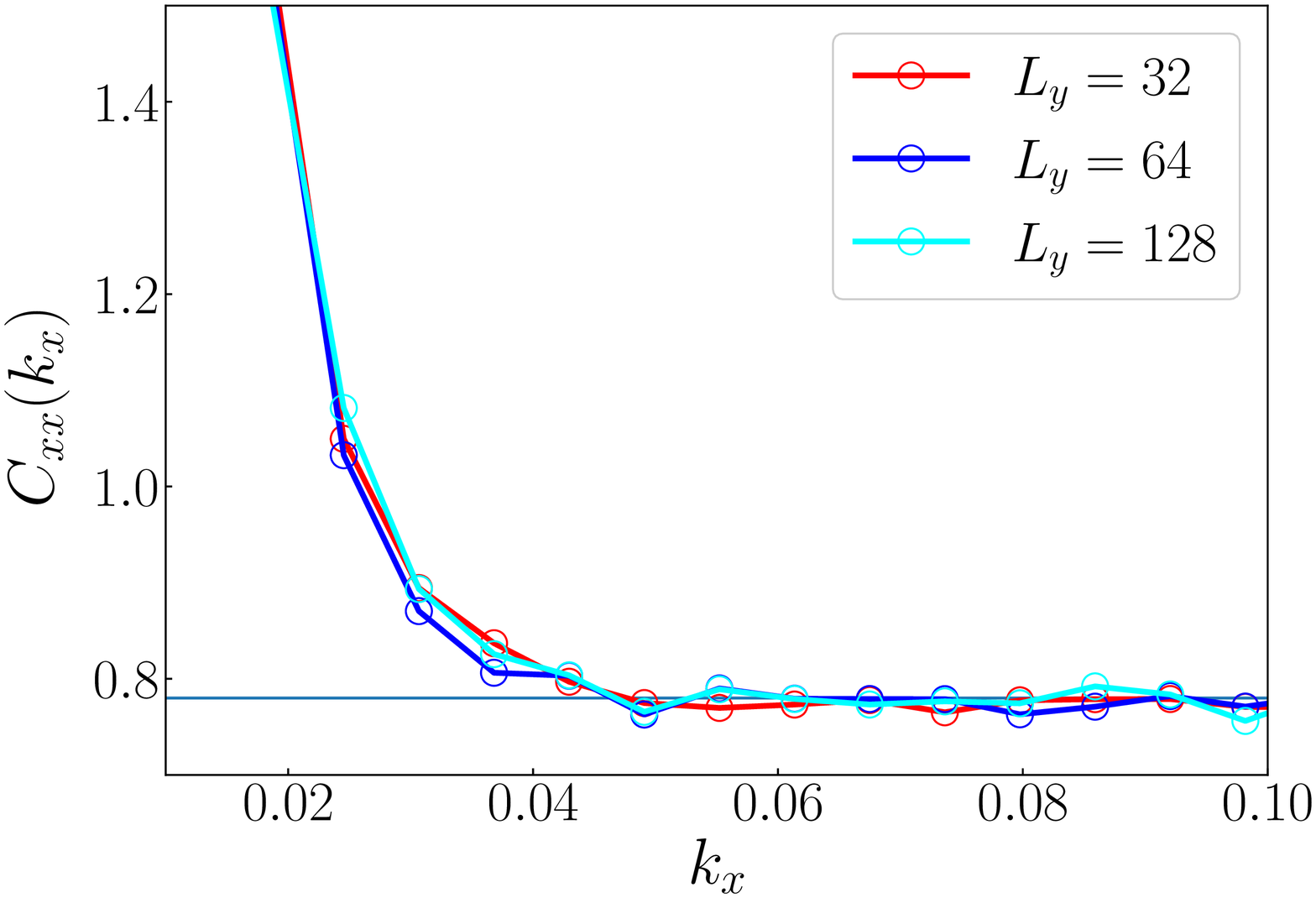} 
\end{center}
\end{minipage}
\begin{minipage}{0.5\hsize}
\begin{center}
\includegraphics[width=4.3cm]{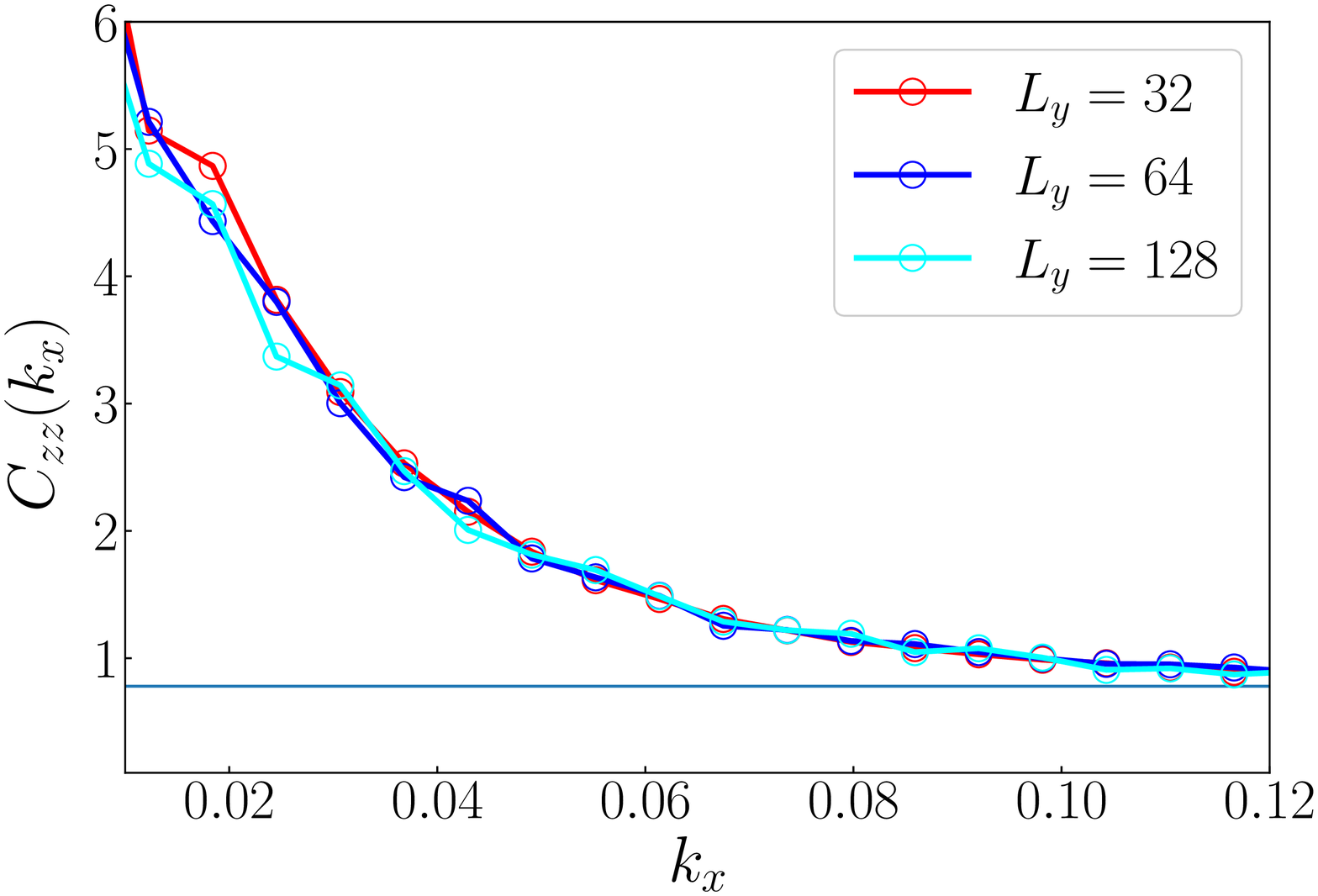} 
\end{center}
\end{minipage}
\end{tabular}
\caption{Left: $L_y$-dependence of $C_{xx}\kx$. Right: $L_y$-dependence of $C_{zz}\kx$. Red, blue, and cyan lines show the MD result with $L_y=32$, $64$, and $128$, respectively. In all cases, $L_x=1024$, $L_z=256$, and $\dot{\gamma}=0.02$.}
\label{fig:MD result: Ly dependence}
\end{figure}
\begin{figure*}[htb]
\centering
\begin{tabular}{cc}
\begin{minipage}{0.5\hsize}
\begin{center}
\includegraphics[width=8.6cm]{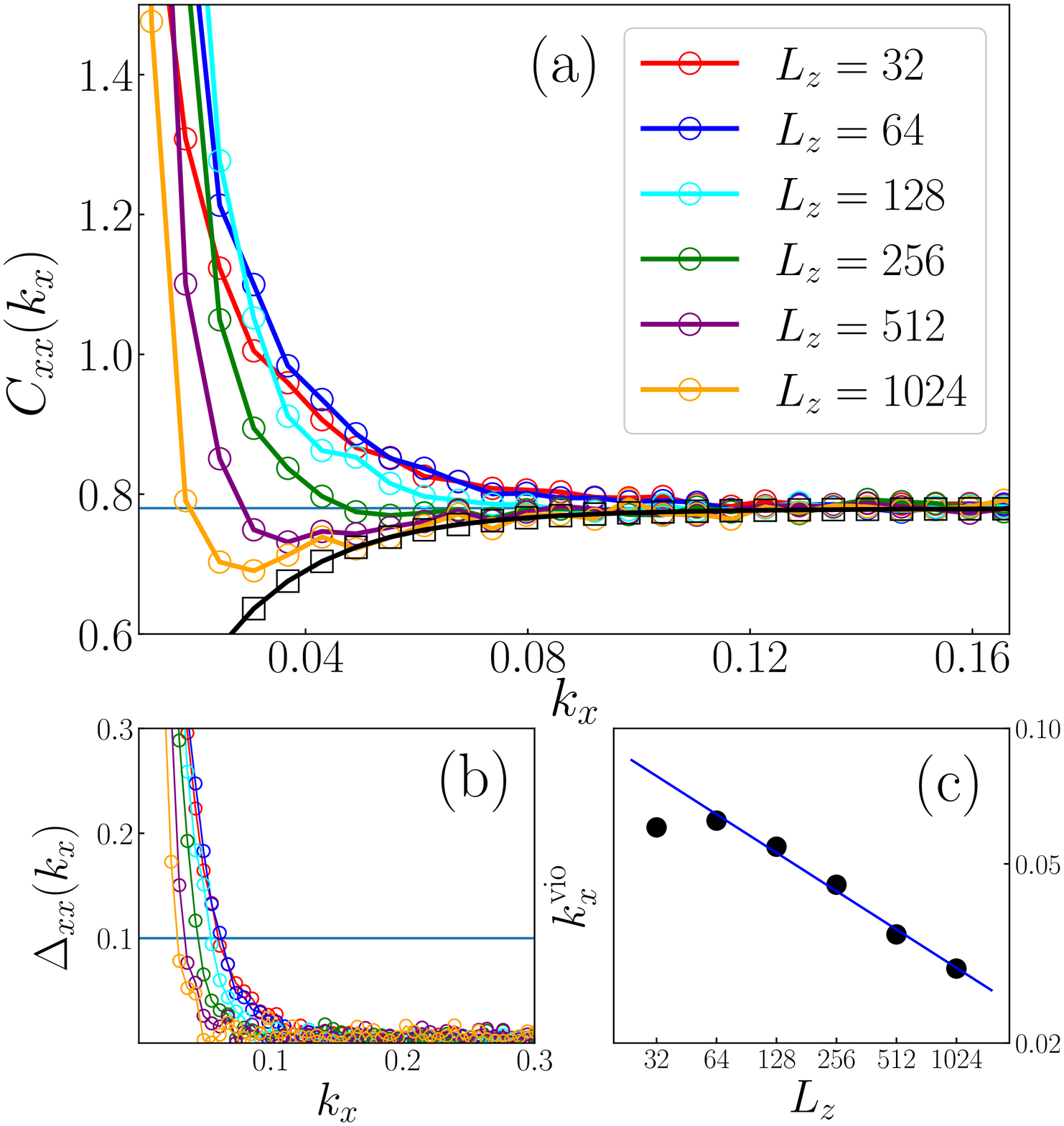} 
\end{center}
\end{minipage}
\begin{minipage}{0.5\hsize}
\begin{center}
\includegraphics[width=8.6cm]{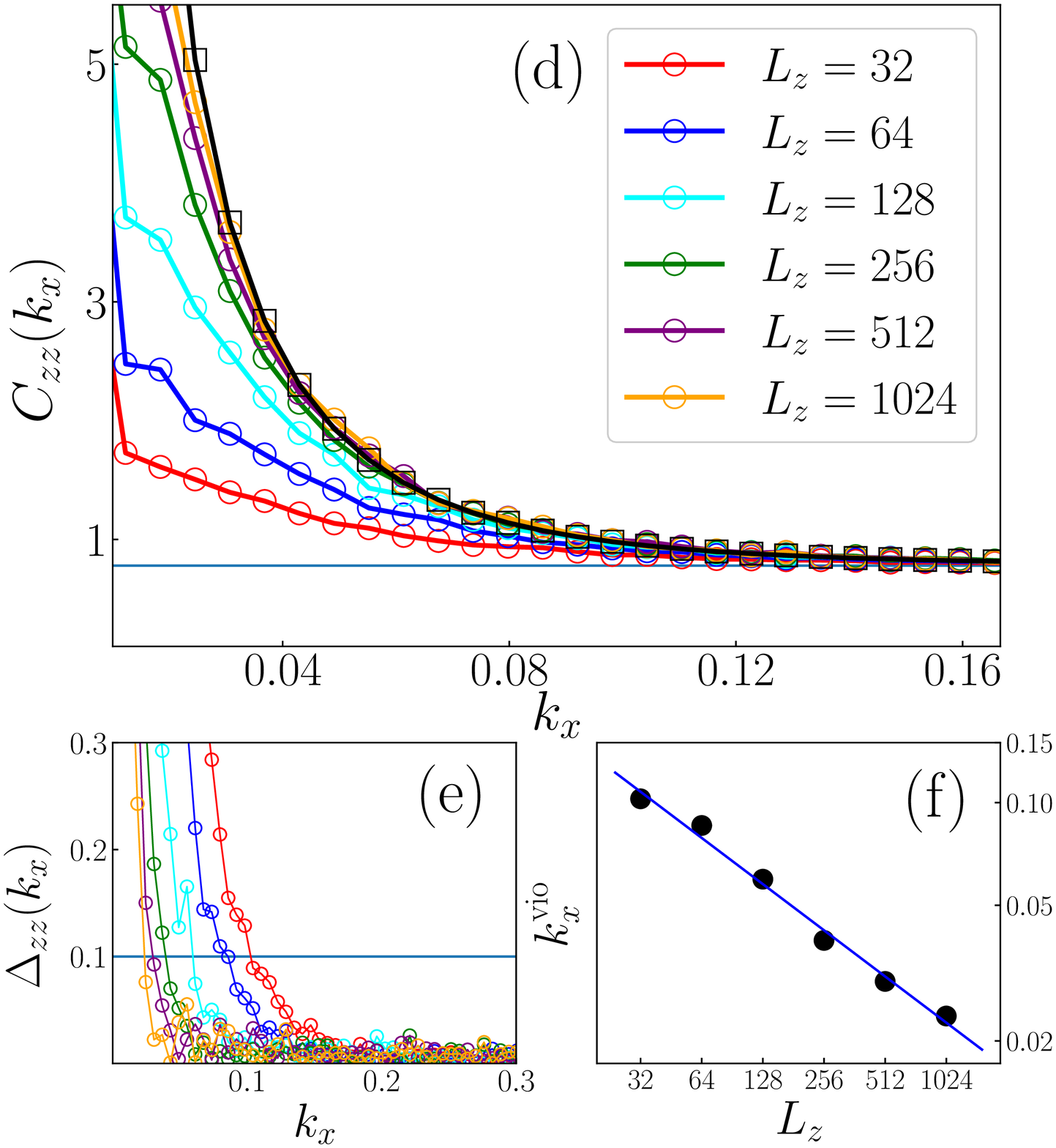} 
\end{center}
\end{minipage}
\end{tabular}
\caption{Left: $L_z$-dependence of $C_{xx}\kx$. (a) $C_{xx}\kx$ for various system sizes. (b) $\Delta_{xx}\kx$ for various system sizes. (c) $k_x^{\rm vio}$ as a function of $L_z$, calculated using the data in (b). Right: $L_z$-dependence of $C_{zz}\kx$. (d) $C_{zz}\kx$ for various system sizes. (e) $\Delta_{zz}\kx$ for various system sizes. (f) $k_x^{\rm vio}$ as a function of $L_z$, calculated using the data in (e). In all cases, $L_x=1024$, $L_y=32$, and $\dot{\gamma}=0.02$. In (a), (b), (d) and (e), the red, blue, cyan, green, purple, and orange lines show the MD result with $L_z=32$, $64$, $128$, $256$, $512$, and $1024$, respectively. The black line shows the LFH solution. The orange line is the same as Fig.~\ref{fig: comparison between MD and LFH: largest system}.}
\label{fig:MD result: Lz dependence}
\end{figure*}
We first examine the MD result for the various system sizes.
Figure~\ref{fig:MD result: Ly dependence} shows that the MD result does not depend on $L_y$. 
In contrast, we can observe strong $L_z$-dependence in Fig.~\ref{fig:MD result: Lz dependence}, where the MD result approaches the LFH solution as $L_z$ increases.

In Figs.~\ref{fig:MD result: Lz dependence}-(a) and -(d), we present the $L_z$-dependence of $C^{\rm MD}_{xx}\kx$ and $C^{\rm MD}_{zz}\kx$.
Figure~\ref{fig:MD result: Lz dependence}-(d) shows that the nonequilibrium LRC of $C^{\rm MD}_{zz}\kx$ gradually grows from the equilibrium value as $L_z$ increases. 
In contrast, Fig.~\ref{fig:MD result: Lz dependence}-(a) shows that the nonequilibrium correction of $C^{\rm MD}_{xx}\kx$ is positive for the small system sizes of $L_z=32$, $64$, and $128$. 
This behavior is inconsistent with the LFH solution; the correction of $C^{\rm FH}_{xx}\kx$, which is the second term of Eq.~(\ref{eq:correlation function: integral expression: xx}), is always negative. 
As $L_z$ increases, the correction dips into the negative region and the positive correction region becomes smaller.
We can infer that the positive correlations for smaller $L_z$ mainly come from finite-size and boundary effects.

We now examine how the MD result approaches the LFH solution as $L_z$ increases.
The deviations $\Delta_{xx}\kx$ and $\Delta_{zz}\kx$ are plotted in Figs.~\ref{fig:MD result: Lz dependence}-(b) and -(e). Moreover, the characteristic wavenumber $k_x^{\rm vio}$ is plotted as a function of $L_z$ in Figs.~\ref{fig:MD result: Lz dependence}-(c) and -(f).
The figures suggest that $k_x^{\rm vio}$ scales as $k_x^{\rm vio} \propto L_z^{-\omega}$ for a fixed $\dot{\gamma}$. By fitting the data with the functional form $k_x^{\rm vio} = A L_z^{-\omega}$, we obtain the following scaling relations with nontrivial exponents:
\begin{eqnarray}
k_x^{\rm vio} = 0.209L_z^{-0.283} \label{kxvio1}
\end{eqnarray}
for $C_{xx}\kx$, and
\begin{eqnarray}
k_x^{\rm vio} = 0.512L_z^{-0.450} \label{kxvio1.5}
\end{eqnarray}
for $C_{zz}\kx$. 
These are depicted by the blue lines in Figs.~\ref{fig:MD result: Lz dependence}-(c) and -(f). Note that Eqs.~(\ref{kxvio1}) and (\ref{kxvio1.5}) are the quantitative relations and enable us to estimate the finite-size effects.

\begin{figure}[thb]
\centering
\begin{tabular}{cc}
\begin{minipage}{0.5\hsize}
\begin{center}
\includegraphics[width=4.3cm]{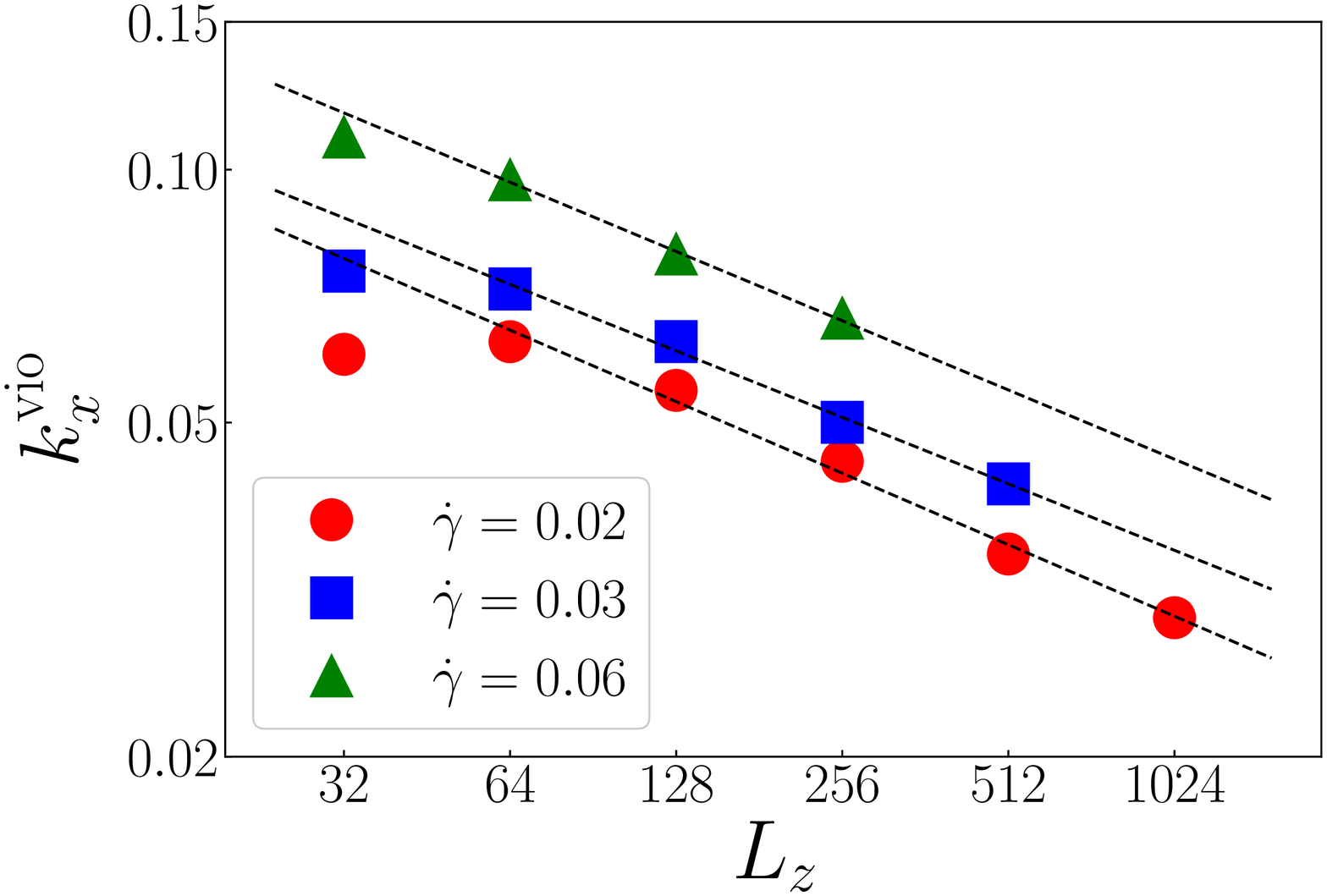} 
\end{center}
\end{minipage}
\begin{minipage}{0.5\hsize}
\begin{center}
\includegraphics[width=4.3cm]{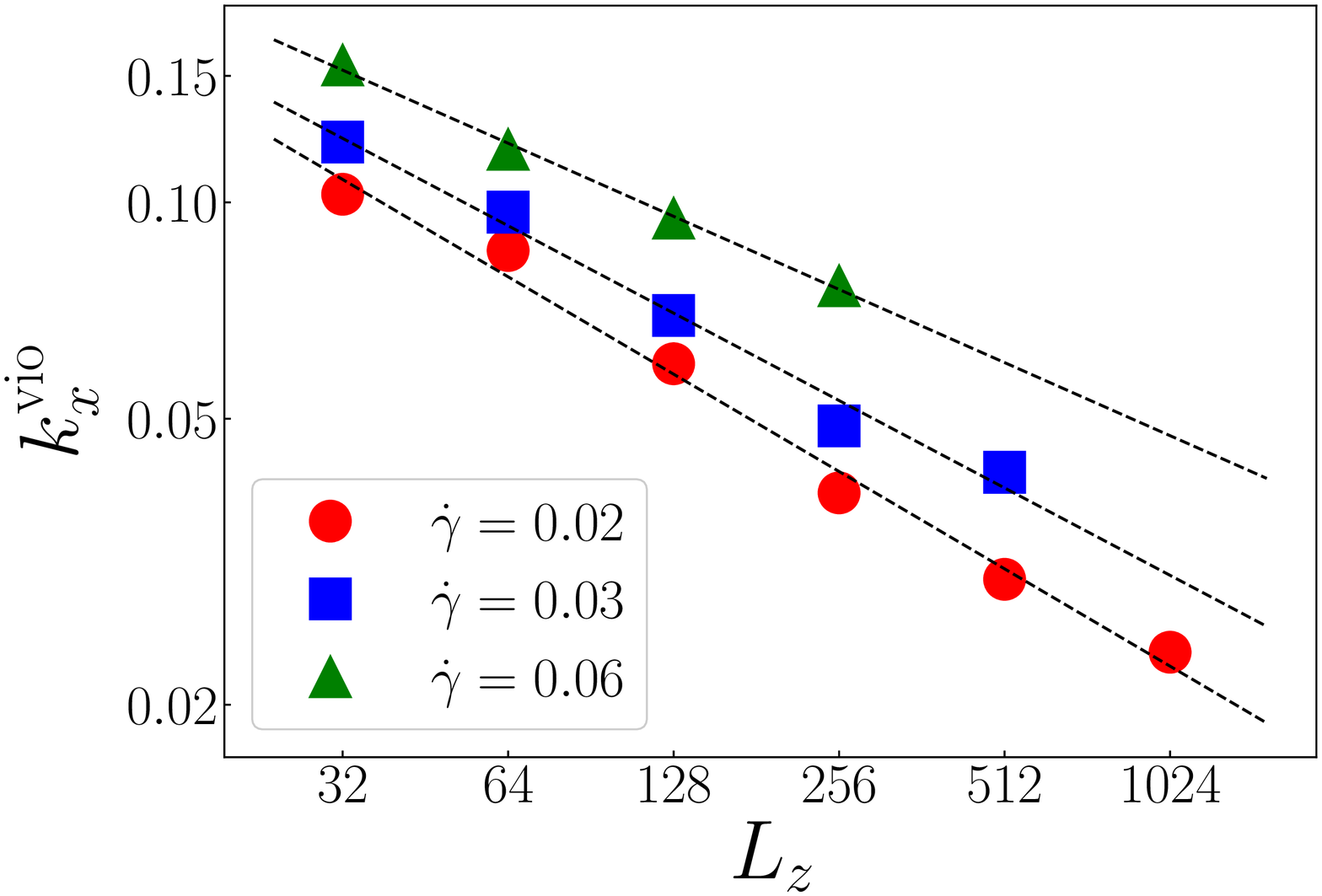} 
\end{center}
\end{minipage}
\end{tabular}
\caption{$k_x^{\rm vio}$ as a function of $L_z$ calculated using the data of $C_{xx}\kx$ (left) and $C_{zz}\kx$ (right) for various shear rates. In all cases, $L_x=1024$ and $L_y=32$. Red, blue, and green symbols show the results for $\dot{\gamma}=0.02$, $0.03$, and $0.06$, respectively. The black lines are the fitting results, which are as follows. Left: $k_x^{\rm vio} = 0.209L_z^{-0.283}$ for $\dot{\gamma}=0.02$, $k_x^{\rm vio} = 0.218L_z^{-0.263}$ for $\dot{\gamma}=0.03$, and $k_x^{\rm vio} = 0.302L_z^{-0.274}$ for $\dot{\gamma}=0.06$. Right: $k_x^{\rm vio} = 0.512L_z^{-0.450}$ for $\dot{\gamma}=0.02$, $k_x^{\rm vio} = 0.498L_z^{-0.404}$ for $\dot{\gamma}=0.03$, and $k_x^{\rm vio} = 0.493L_z^{-0.338}$ for $\dot{\gamma}=0.06$.}
\label{fig:MD result: summary of kviolation}
\end{figure}

Furthermore, we consider the dependence of the scaling relation on the shear rate $\dot{\gamma}$.
We plot $k_x^{\rm vio}$ as a function of $L_z$ for several $\dot{\gamma}$ in Fig.~\ref{fig:MD result: summary of kviolation}.
The figure shows that the scaling form $k_x^{\rm vio} \propto L_z^{-\omega}$ holds, regardless of the value of $\dot{\gamma}$.
For $C_{zz}\kx$, $\omega$ takes values of $0.34$--$0.45$ depending on $\dot{\gamma}$. 
In contrast, for $C_{xx}\kx$, $\omega$ is close to $0.27$, and is largely insensitive to $\dot{\gamma}$.

\subsection{Shear-rate dependence of nonequilibrium LRC}
\begin{figure*}[thb]
\centering
\begin{tabular}{cc}
\begin{minipage}{0.6\hsize}
\begin{center}
\includegraphics[width=10.56cm]{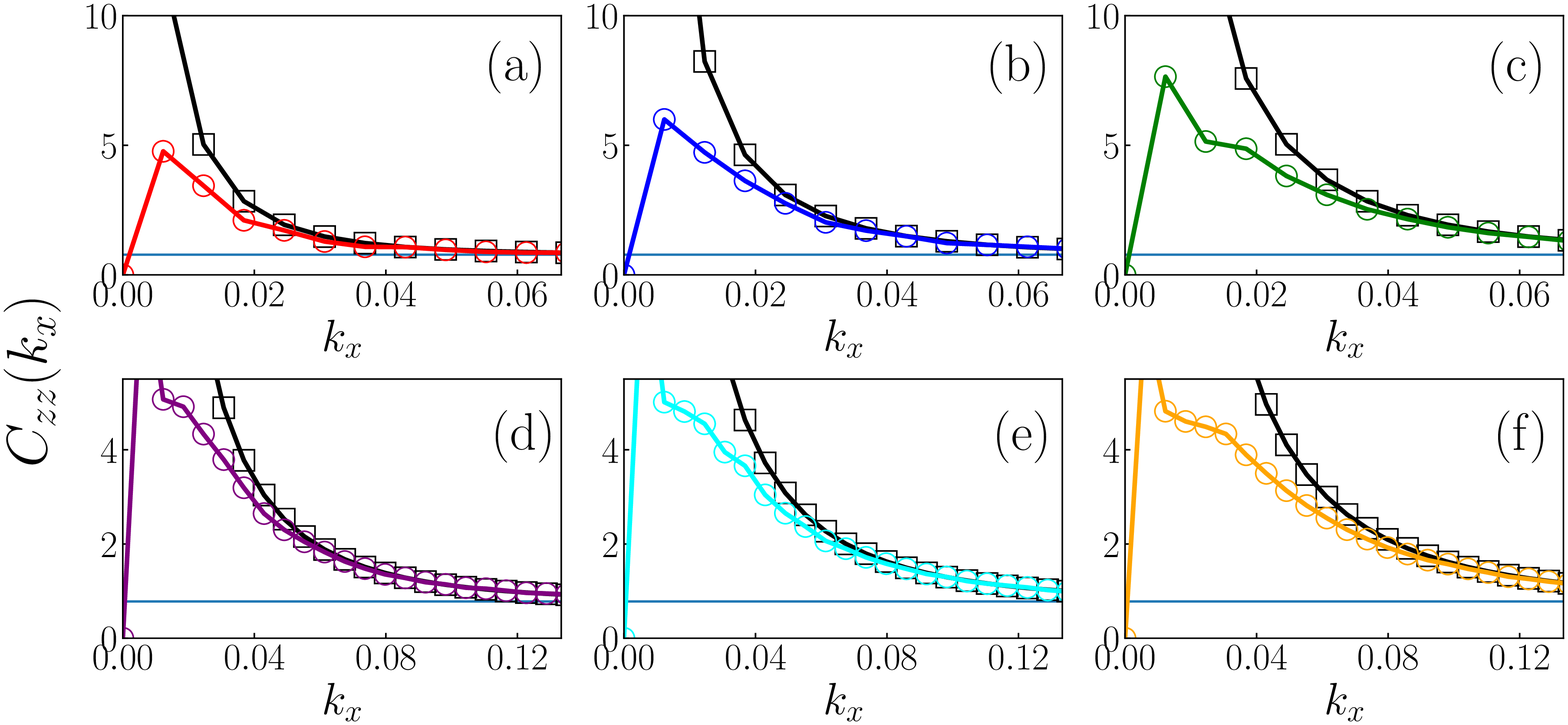} 
\end{center}
\end{minipage}
\begin{minipage}{0.4\hsize}
\begin{center}
\includegraphics[width=7.04cm]{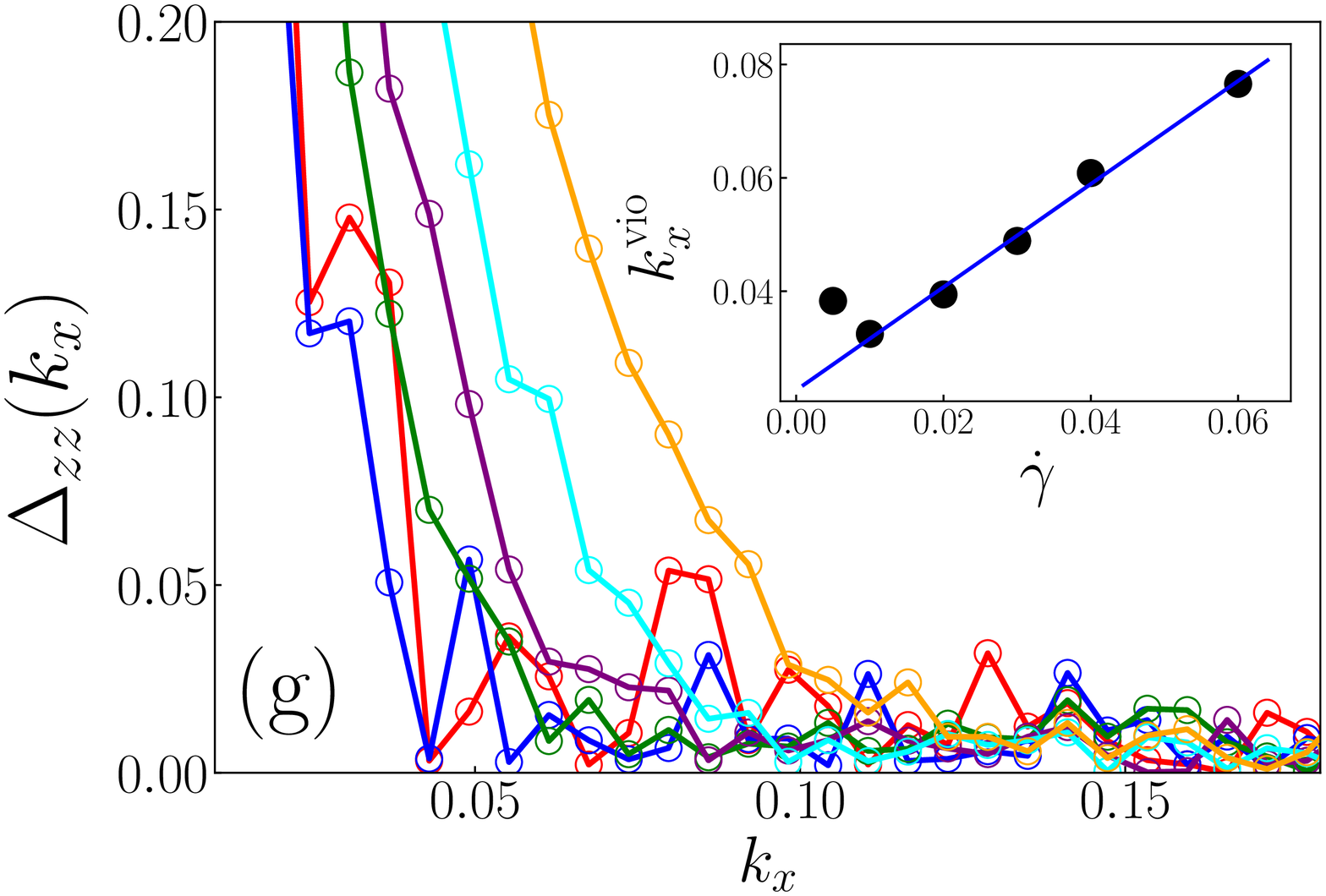} 
\end{center}
\end{minipage}
\end{tabular}
\caption{(a)--(f): $\dot{\gamma}$-dependence of $C_{zz}\kx$ for various shear rates. (g): $\Delta_{zz}\kx$ as a function of $k_x$. The system size is $L_x=1024, L_y=32, L_z=256$. (a) $\dot{\gamma}=0.005$, (b) $\dot{\gamma}=0.01$, (c) $\dot{\gamma}=0.02$, (d) $\dot{\gamma}=0.03$, (e) $\dot{\gamma}=0.04$, (f) $\dot{\gamma}=0.06$. The color in (g) corresponds to that in (a)--(f). Inset in (g) shows $k^{\rm vio}_x$ as a function of $\dot{\gamma}$. The blue line is the fitting result, which is given by $k_x^{\rm vio} = 0.924\dot{\gamma}+0.011$.}
\label{fig:MD result: srate dependence}
\end{figure*}
We now examine the shear-rate dependence of the LRCs for a fixed system size.
In Fig.~\ref{fig:MD result: srate dependence}, we plot the correlation $C_{zz}\kx$ and the deviation $\Delta_{zz}\kx$ for various values of $\dot{\gamma}$ from $0.005$ to $0.06$.
We observe that the deviation increases monotonically as $\dot{\gamma}$ increases from $0.01$ to $0.06$ in Fig.~\ref{fig:MD result: srate dependence}-(g).
However, the result for $\dot{\gamma}=0.005$ does not exhibit this tendency.  
We can infer that the LRC does not fully develop when $\dot{\gamma}=0.005$ because $L_z$ is too small.

The inset of Fig.~\ref{fig:MD result: srate dependence}-(g) shows $k_x^{\rm vio}$ as a function of $\dot{\gamma}$. Clearly, $k_x^{\rm vio}$ is linearly dependent on $\dot{\gamma}$ from $\dot{\gamma}=0.01$ to $\dot{\gamma}=0.06$. 
By fitting this with the functional form $A\dot{\gamma} + B$, we obtain the quantitative relation
\begin{eqnarray}
k_x^{\rm vio} = 0.924\dot{\gamma}+0.011.
\end{eqnarray}
Similar behavior can be observed for $C_{xx}\kx$,
\begin{eqnarray}
k_x^{\rm vio}=0.583\dot{\gamma}+0.032.\label{kxvio2}
\end{eqnarray}

\subsection{Kinetic temperature and pressure}
\begin{figure}[htb]
\centering
\begin{tabular}{cc}
\begin{minipage}{0.5\hsize}
\begin{center}
\includegraphics[width=4.3cm]{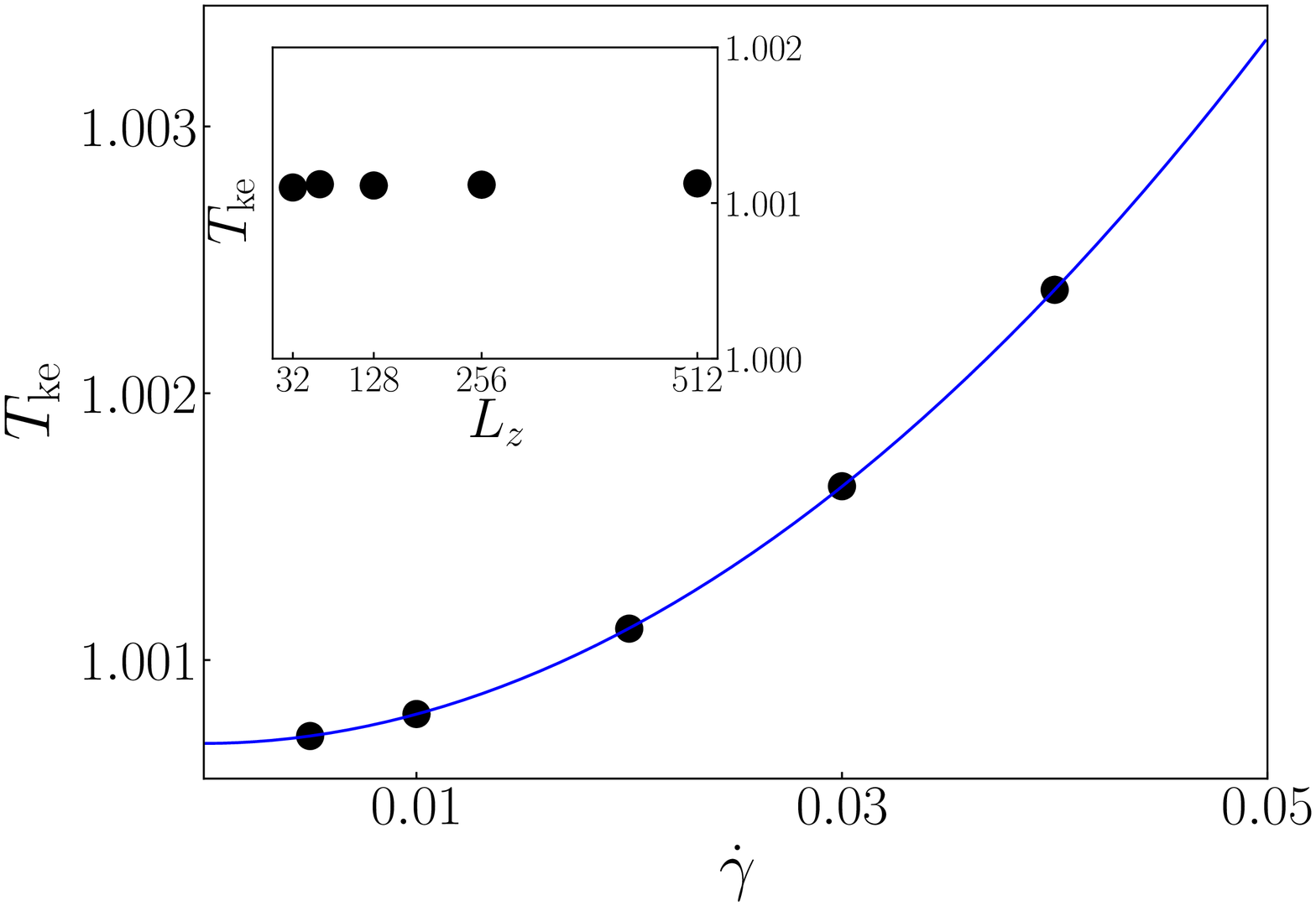} 
\end{center}
\end{minipage}
\begin{minipage}{0.5\hsize}
\begin{center}
\includegraphics[width=4.3cm]{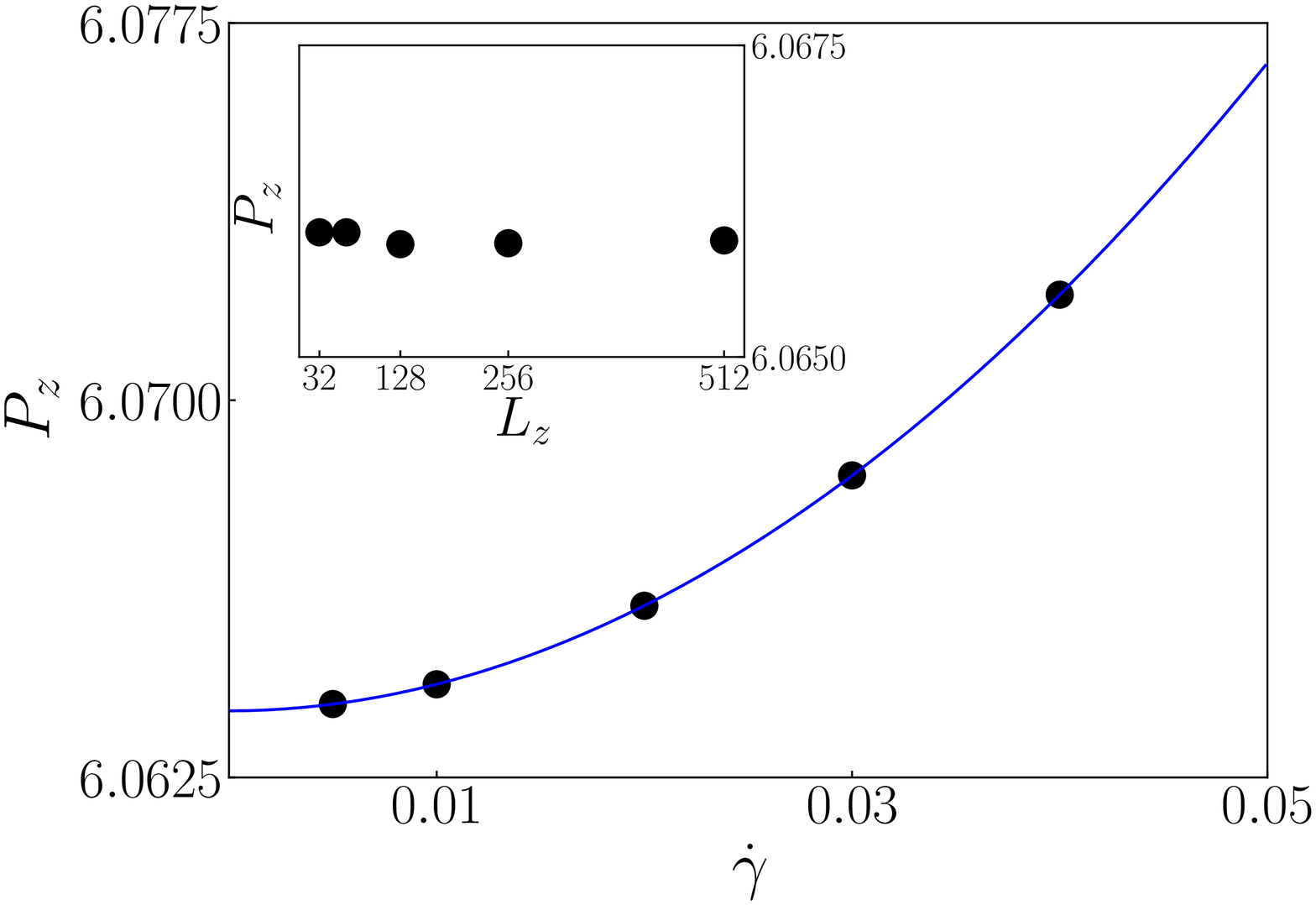} 
\end{center}
\end{minipage}
\end{tabular}
\caption{Left: kinetic temperature $T_{\rm ke}$ as a function of $\dot{\gamma}$ for $L_x=1024$, $L_y=32$, and $L_z=256$. The blue line is the fitting result given by $0.9894\dot{\gamma}^{1.9777}+ 1.00069$. Inset shows the $L_z$-dependence of the kinetic temperature $T_{\rm ke}$ for $\dot{\gamma}=0.02$. Right: pressure $P^z$ as a function of $\dot{\gamma}$ for $L_x=1024$, $L_y=32$, and $L_z=256$. The blue line is the fitting result given by $4.8985\dot{\gamma}^{1.9831}+6.0638$. Inset shows the $L_z$-dependence of the pressure $P^z$ for $\dot{\gamma}=0.02$. }
\label{fig:MD result: kinetic temperature and pressure}
\end{figure}

Finally, we study the kinetic temperature $T_{\rm ke}$, which is defined as
\begin{eqnarray}
T_{\rm ke} \equiv \frac{1}{2N} \Big\langle \sum_{i=1}^N \Big(\frac{p_{iy}^2}{m}+ \frac{p_{iz}^2}{m} \Big)\Big\rangle ,
\end{eqnarray}
and the pressure $P_{z}$ along the $z$-direction, which is defined as
\begin{eqnarray}
P_{z} &=& \frac{1}{V} \Big\langle\sum_{i=1}^N \frac{p^2_{iz}}{m} + \sum_{i=1}^N \sum_{j>i} (z_{i}-z_{j})  f_{ij, z}\Big\rangle.
\end{eqnarray}
Here, $ f_{ij, z}$ is the $z$-component of the intermolecular force between particles $i$ and $j$.
These quantities are plotted in Fig.~\ref{fig:MD result: kinetic temperature and pressure}. 
Previous MD simulations~\cite{Naitoh1978the,Naitoh1979shear,Hoover1980lennard,Evans1989on,Evans1990viscosity,Travis1998strain,Istvan2002shear,Evans1980computer,Erpenbeck1984shear,Marcelli2001analytic,Jlian2001energy,Ahmed2009strain,Jialin2003scaing,Todd2005power,Desgranges2009universal,Lautenschlaeger2019shear} have explored the $\dot{\gamma}$-dependence of the kinetic temperature and pressure to probe the LRC, as explained in the Introduction. 
Following these studies, we fit the simulation data to the form $A\dot{\gamma}^B+C$ and obtain
\begin{eqnarray}
T_{\rm ke} &=& 0.9894\dot{\gamma}^{1.9777}+ 1.00069,
\label{eq:fitting results of kinetic temperature}\\ [3pt]
P_z &=& 4.8985\dot{\gamma}^{1.9831}+6.0638.
\label{eq:fitting results of pressure}
\end{eqnarray}
Both exponents are close to $2$. 

We now consider whether the long- or short-range contributions dominate our result, as suggested in Ref.~\cite{DeZarate2019nonequilibrium}.
To this end, we decompose the corrections into the contributions from the short- and the long-range scales:
\begin{eqnarray}
\delta T_{\rm ke} = \delta T_{\rm ke}^{\rm SR} + \delta T_{\rm ke}^{\rm LR}, \\ [3pt]
\delta P_{z} = \delta P_{z}^{\rm SR} + \delta P_{z}^{\rm LR}.
\end{eqnarray}
Our result suggests the dominance of the short-range scale as follows. 
The correction proportional to $\dot{\gamma}^2$ in the fluctuating hydrodynamics is linearly dependent on the system size $L$:
\begin{eqnarray}
\delta T_{\rm ke}^{\rm LR} \sim \delta P_{z}^{\rm LR} \sim L\dot{\gamma}^{2},
\end{eqnarray}
which comes from the $k_x^{-4}$-behavior of Eq.~(\ref{eq:asymptotic expression for Czz1}). 
However, our result is almost independent of $L_z$, as shown in the inset of Fig.~\ref{fig:MD result: kinetic temperature and pressure}, although our result catches the $k_x^{-4}$-tail. 
Thus, our result supports the assertion that the $\dot{\gamma}^2$-dependence comes from the short-range scale instead of the nonequilibrium LRC. 
However, recall that the LFH solution is not valid in the long-wavelength region, as shown in Fig.~\ref{fig:MD result: Lz dependence}. 
Therefore, further theoretical studies on the short-range corrections are required to form a final conclusion.

\section{Concluding remarks and discussion}
\label{sec:Discussion}
\begin{table*}[t]
\centering
\begin{tabularx}{170mm}{c||C|C|C}
previous study & Otsuki and Hayakawa (2009) & Varghese et al. (2015) &  Our study  \\ \hline \hline
\multirow{2}{*}{model} & hard sphere with & multiparticle collision & Weeks--Chandler--\\
 & restitution coefficient $e$ & dynamics fluid & Andersen fluid \\ \hline
boundary condition & Lees--Edwards & Lees--Edwards & Lees--Edwards \\ \hline
\multirow{2}{*}{thermostat} & \hspace{-1.55cm} $e\neq 1$:  none & cell-level Maxwell--Boltzmann & dissipative particle \\
 & $e=1$: velocity scaling & \hspace{-0.35cm} rescaling of relative velocity & dynamics  \\ \hline
local momentum & $e\neq 1$: yes & \multirow{2}{*}{yes} & \multirow{2}{*}{yes} \\
conservation & $e=1$: no &  &  \\ \hline
typical system size & $32\sigma \sim 112\sigma$ & $20a \sim120a$ & $32\sigma \sim 1024\sigma$ \\ \hline
\end{tabularx}
\caption{Setup of simulations in the previous studies. $\sigma$ and $a$ are, respectively, typical length scales characterizing the particle size or interaction range.}
\label{tab: setup of simulations in the previous studies}
\end{table*}
Let us compare our result to those reporeted by Otsuki and Hayakawa~\cite{Otsuki2009spatial} and Varghese et al.~\cite{Varghese2017spatial}. 
These previous studies directly observed the shear-induced LRC in particle-based simulations. 
Table~\ref{tab: setup of simulations in the previous studies} presents a comparison of their setups with that of our simulations. 
The system size in our study is about $10$ times larger than that in the previous studies. As a result, we can systematically study the finite-size effect on shear-induced LRCs. 
We showed that the LFH solution is quantitatively consistent with the MD result when $L_z$ is sufficiently large. 
Conversely, for smaller $L_z$, the MD result deviates from the LFH solution in the long-wavelength region. Such deviations were also observed in the previous studies~\cite{Otsuki2009spatial,Varghese2017spatial}. 
Varghese et al.~proposed that these deviations originated from the density-dependence of viscosity. 
However, our simulations have clarified that the derivations are caused by an insufficient system size. 

Furthermore, we examined how the deviations depend on the system size and the shear rate.
As a quantitative examination, we introduced the characteristic wavenumber $k_x^{\rm vio}$ associated with the breakdown of the hydrodynamic description.
$k_x^{\rm vio}$ determines the applicable wavenumber region of the LFH solution as  $k_x>k_x^{\rm vio}$.
We then found two scaling relations, $k_x^{\rm vio} \propto L_z^{-\omega}$ at fixed $\dot{\gamma}$ and $k_x^{\rm vio} \propto \dot{\gamma}$, for a fixed $L_z$.

The interesting point is that the finite-size effect is non-negligible in a large region. 
For example, $k_x^{\rm vio}$ is obtained from Fig.~\ref{fig: comparison between MD and LFH: largest system} as $k_x^{\rm vio}\simeq 0.0237$ for $L_z=1024$ and $\dot{\gamma}=0.02$. 
In the real space, the corresponding $x^{\rm vio}\equiv 2\pi/k_x^{\rm vio}$ is about $265$. Therefore, if we consider the system with $L_x=L_y=L_z=1024$, the LFH solution breaks down in about three-quarters of the region, $0.26L_x < x < L_x$, where large finite-size effects exist.
Note that the magnitude of the finite-size effects is related to the value of the exponent $\omega$. The scaling relation $k_x^{\rm vio} \propto L_z^{-\omega}$ can be rewritten as $x^{\rm vio} \propto L_z^{\omega}$. By noting that the breakdown of the LFH solution occurs in the region $L_z^{\omega} < x < L_x$, we find that a smaller $\omega$ yields finite-size effects in a larger region. Actually, $\omega = 0.27$--$0.45$ as our model is quite small.

The question to be asked is the origin of such large finite-size effects. 
We can infer that they come from the Lees--Edwards boundary condition and the nonlinearity of the fluctuating hydrodynamics. Future work should analyze these effects in the fluctuating hydrodynamics. As a related problem, it is interesting how the hydrodynamic description predicts the exponent $\omega$.

Finally, we remark on the utility of the quantitative relations for $k_x^{\rm vio}$, such as Eqs.~(\ref{kxvio1})--(\ref{kxvio2}). They enable us to estimate the finite-size effects in larger-size simulations from smaller-size simulations. 
For example, we can use the estimation to observe the $k_x^{-4/3}$-tail of $C_{zz}(k_x)$.
We could not observe this tail as shown in the right-hand panel of Fig.~\ref{fig: comparison between MD and LFH: largest system} because the hydrodynamic description breaks down before $C_{zz}(k_x)$ exhibits the $k_x^{-4/3}$-tail.
To observe the $k_x^{-4/3}$-tail at $\dot{\gamma}=0.02$, we need to reduce $k_x^{\rm vio}$ to $0.01$. This value is estimated from the LFH solution. The required $L_z$ is then calculated from Eq.~(\ref{kxvio1}) as $L_z\simeq 6286$. Such a quantitative estimation is useful for preparing larger-size simulations and laboratory experiments.

\bigskip
\sectionprl{Acknowledgements}
We thank Hiroshi Watanabe, Shi-ichi Sasa, and Naoko Nakagawa for the fruitful discussions. The computations in this study were performed using the facilities of the Supercomputer Center at the Institute for Solid State Physics, The University of Tokyo. H.N. is supported by KAKENHI Grant Number JP21J00034. Y.M. is supported by the Zhejiang Provincial Natural Science Foundation Key Project (Grant No. LZ19A050001) and NSF of China (Grants Nos. 11975199 and 11674283).

\bigskip
\appendix
\section{Brief sketch of derivation of Eqs.~(\ref{eq:correlation function: integral expression: xx}) and (\ref{eq:correlation function: integral expression: zz}) }
\label{appendix: Brief sketch of derivation}
To derive the integral expressions in Eqs.~(\ref{eq:correlation function: integral expression: xx}) and (\ref{eq:correlation function: integral expression: zz}) for $C_{xx}\kx$ and $C_{zz}\kx$, we use two approximations. Here, we briefly sketch their derivation while focusing on these approximations.

The first approximation is to neglect the nonlinear fluctuations. $\rho(\bm{r},t)$, $p(\bm{r},t)$, and $\bm{v}(\bm{r},t)$ are expanded around the zero-order solution as
\begin{eqnarray}
\rho(\bm{r},t) &=& \rho_0 + \delta\rho(\bm{r},t), \nonumber \\[3pt]
p(\bm{r},t) &=& p_0 + c_T^2 \delta \rho, \nonumber \\[3pt]
v_x(\bm{r},t) &=& \dot{\gamma} z + \delta v_x(\bm{r},t),
\label{eq:first order solution} \\[3pt]
v_y(\bm{r},t) &=& \delta v_y(\bm{r},t), \nonumber \\[3pt]
v_z(\bm{r},t) &=& \delta v_z(\bm{r},t),\nonumber
\end{eqnarray} 
where $c_T$ is the isothermal speed of sound. By substituting Eq.~(\ref{eq:first order solution}) into Eqs.~(\ref{eq: basic EOM0}) and (\ref{eq: basic EOM1}) and neglecting the higher-order terms of $\delta \rho$ and $\delta \bm{v}$, we have
\begin{widetext}
\begin{eqnarray}
& & \Biggl(\frac{\partial}{\partial t} - \dot{\gamma} k_x \frac{\partial}{\partial k_z} \Biggr)
\begin{pmatrix}
\delta \tilde{\rho} \\
\delta \tilde{v}_x \\
\delta \tilde{v}_y \\
\delta \tilde{v}_z \\
\end{pmatrix}
+
\begin{pmatrix}
0 & 0 & 0 & 0\\
0 & 0 & 0 & \dot{\gamma}\\
0 & 0 & 0 & 0\\
0 & 0 & 0 & 0\\
\end{pmatrix} 
\begin{pmatrix}
\delta \tilde{\rho} \\
\delta \tilde{v}_x \\
\delta \tilde{v}_y \\
\delta \tilde{v}_z \\
\end{pmatrix}
\nonumber \\[3pt]
& & = -
\begin{pmatrix}
0 & -ik_x & -ik_y & -ik_z \\
-i \frac{c_T^2}{\rho_0} k_x & \frac{\eta}{\rho_0} |\bm{k}|^2 + \frac{\eta^k}{3\rho_0} k_x^2 & \frac{\eta^k}{3\rho_0} k_x k_y & \frac{\eta^k}{3\rho_0} k_x k_z \\
-i \frac{c_T^2}{\rho_0} k_y & \frac{\eta^k}{3\rho_0} k_x k_y & \frac{\eta}{\rho_0} |\bm{k}|^2 + \frac{\eta^k}{3\rho_0} k_y^2 & \frac{\eta^k}{3\rho_0} k_y k_z \\
-i \frac{c_T^2}{\rho_0} k_z & \frac{\eta^k}{3\rho_0} k_x k_z & \frac{\eta^k}{3\rho_0} k_y k_z & \frac{\eta}{\rho_0} |\bm{k}|^2 + \frac{\eta^k}{3\rho_0} k_z^2 \\
\end{pmatrix} 
\begin{pmatrix}
\delta \tilde{\rho} \\
\delta \tilde{v}_x \\
\delta \tilde{v}_y \\
\delta \tilde{v}_z \\
\end{pmatrix}
-
\begin{pmatrix}
0 \\
i k_j \tilde{s}_{xj} \\
i k_j \tilde{s}_{yj} \\
i k_j \tilde{s}_{zj} \\
\end{pmatrix}.
\label{eq:linear equation of motion in the Fourier space}
\end{eqnarray}
\end{widetext}

Under this approximation, the momentum correlations are connected with the velocity correlations as
\begin{eqnarray}
\langle \delta g_i(\bm{r},t) \delta g_j(\bm{r}',t) \rangle &\simeq& \rho_0^2\langle \delta v_i(\bm{r},t) \delta v_j(\bm{r}',t) \rangle .
\label{eq: relation between momentum and velocity field}
\end{eqnarray}

The second approximation is used when we decompose the longitudinal and transverse waves. Here, it is convenient to introduce the oblique coordinate used by Lutsko and Dufty~\cite{Lutsko1985hydrodynamic}:
\begin{eqnarray}
\begin{pmatrix}
\tilde{\xi}_1(\bm{k},t) \\
\tilde{\xi}_2(\bm{k},t) \\
\tilde{\xi}_3(\bm{k},t) \\
\tilde{\xi}_4(\bm{k},t) \\
\end{pmatrix}
=
\begin{pmatrix}
(c_T/\rho_0) \delta \tilde{\rho}(\bm{k},t) \\
\delta \tilde{\bm{v}}(\bm{k},t) \cdot \hat{\bm{e}}^{\rm (1)}(\bm{k}) \\
\delta \tilde{\bm{v}}(\bm{k},t) \cdot \hat{\bm{e}}^{\rm (2)}(\bm{k}) \\
\delta \tilde{\bm{v}}(\bm{k},t) \cdot \hat{\bm{e}}^{\rm (3)}(\bm{k}) \\
\end{pmatrix}.
\label{eq:transformation into oblique coordinate}
\end{eqnarray}
The vectors $\{\hat{\bm{e}}^{\rm (a)}(\bm{k})\}_{a=1,2,3}$ are a set of orthogonal unit vectors given by
\begin{eqnarray}
\hat{\bm{e}}^{\rm (1)}(\bm{k}) = \hat{\bm{k}} = 
\begin{pmatrix}
\frac{k_x}{k}\\
\frac{k_y}{k} \\
\frac{k_z}{k} \\
\end{pmatrix},
\end{eqnarray}
\begin{eqnarray}
\hat{\bm{e}}^{\rm (2)}(\bm{k}) = \frac{\hat{\bm{z}} - \hat{\bm{e}}^{\rm (1)}(\bm{k})(\hat{\bm{e}}^{\rm (1)}(\bm{k})\cdot \hat{\bm{z}} )}{\hat{k}_{\perp}} =
\begin{pmatrix}
- \frac{k_x k_z}{k k_{\perp}} \\
- \frac{k_y k_z}{k k_{\perp}} \\
\hat{k}_{\perp} \\
\end{pmatrix},
\end{eqnarray}
\begin{eqnarray}
\hat{\bm{e}}^{\rm (3)}(\bm{k}) = \hat{\bm{e}}^{\rm (1)}(\bm{k}) \times \hat{\bm{e}}^{\rm (2)}(\bm{k}) = 
\begin{pmatrix}
\frac{k_y}{k_{\perp}} \\
- \frac{k_x}{k_{\perp}} \\
0 \\
\end{pmatrix},
\end{eqnarray}
where $k = |\bm{k}|$, $k_{\perp} = \sqrt{k^2-k_z^2}$, and $\hat{k}_{\perp} = \sqrt{k^2-k_z^2}/k$.

The time evolution of $\tilde{\bm{\xi}}(\bm{k},t)$ is immediately obtained by substituting the inverse transformation of Eq.~(\ref{eq:transformation into oblique coordinate}) into Eq.~(\ref{eq:linear equation of motion in the Fourier space}) as
\begin{eqnarray}
& & \Biggl(\frac{\partial}{\partial t} - \dot{\gamma} k_x \frac{\partial}{\partial k_z} \Biggr)
\begin{pmatrix}
\tilde{\xi}_1 \\
\tilde{\xi}_2 \\
\tilde{\xi}_3 \\
\tilde{\xi}_4 \\
\end{pmatrix}
+ L(\bm{k})
\begin{pmatrix}
\tilde{\xi}_1 \\
\tilde{\xi}_2 \\
\tilde{\xi}_3 \\
\tilde{\xi}_4 \\
\end{pmatrix}
= 
\begin{pmatrix}
\tilde{f}_1 \\
\tilde{f}_2 \\
\tilde{f}_3 \\
\tilde{f}_4 \\
\end{pmatrix}
\label{eq:master equation of xi}
\end{eqnarray}
with
\begin{eqnarray}
L(\bm{k}) = -ik B + k^2 C + \dot{\gamma} D(\bm{k}),
\end{eqnarray}
and
\begin{eqnarray}
\tilde{f}_{1} &=& 0,\\
\tilde{f}_{\alpha+1}&=&\hat{\bm{e}}^{\rm (\alpha)}(\bm{k}) \cdot
\begin{pmatrix}
0 \\
i k_j \tilde{s}_{xj} \\
i k_j \tilde{s}_{yj} \\
i k_j \tilde{s}_{zj} \\
\end{pmatrix},
\end{eqnarray}
where $\alpha=1,2,3$ and the matrices $B$, $C$, and $D(\bm{k})$ are given by
\begin{eqnarray}
B = 
\begin{pmatrix}
0 & c_T & 0 & 0 \\
c_T & 0 & 0 & 0 \\
0 & 0 & 0 & 0 \\
0 & 0 & 0 & 0 \\
\end{pmatrix} ,
\end{eqnarray}
\begin{eqnarray}
C = \frac{1}{\rho_0}
\begin{pmatrix}
0 & 0 & 0 & 0 \\
0 & \eta + \frac{\eta^k}{3} & 0 & 0 \\
0 & 0 & \eta & 0 \\
0 & 0 & 0 & \eta \\
\end{pmatrix} ,
\end{eqnarray}
\begin{eqnarray}
D = 
\begin{pmatrix}
0 & 0 & 0 & 0 \\
0 & \frac{k_x k_z}{k^2} & 2\frac{k_x k_{\perp}}{k^2} & 0 \\
0 & - \frac{k_x}{k_{\perp}} & - \frac{k_x k_z}{k^2} & 0 \\
0 & \frac{k_y k_z}{k k_{\perp}} & \frac{k_y}{k} & 0 \\
\end{pmatrix} .
\end{eqnarray}
To decompose the longitudinal and transverse waves in Eq.~(\ref{eq:master equation of xi}), we need to solve the eigenvalue problem:
\begin{eqnarray}
\hspace{-1cm} \Big(-\dot{\gamma} k_x \frac{\partial}{\partial k_z} + L(\bm{k})\Big)
\begin{pmatrix}
\tilde{\zeta}^{\rm (i)}_1(\bm{k}) \\
\tilde{\zeta}^{\rm (i)}_2(\bm{k}) \\
\tilde{\zeta}^{\rm (i)}_3(\bm{k}) \\
\tilde{\zeta}^{\rm (i)}_4(\bm{k}) \\
\end{pmatrix}
= \lambda_i(\bm{k})
\begin{pmatrix}
\tilde{\zeta}^{\rm (i)}_1(\bm{k}) \\
\tilde{\zeta}^{\rm (i)}_2(\bm{k}) \\
\tilde{\zeta}^{\rm (i)}_3(\bm{k}) \\
\tilde{\zeta}^{\rm (i)}_4(\bm{k}) \\
\end{pmatrix},
\label{eq:diagonalization problem}
\end{eqnarray}
where $\tilde{\bm{\zeta}}^{\rm (i)}(\bm{k})$ and $\lambda_i(\bm{k})$, respectively, are the $i$th eigenvector and eigenvalue ($i=1,2,3,4$).
At zero shear rate, the eigenvalue problem in Eq.~(\ref{eq:diagonalization problem}) reduces to the diagonalization of the matrix $-ikB+k^2C$, which can be solved exactly. However, for a finite shear rate, the longitudinal and transverse waves obtained at zero shear rate are strongly coupled and, as a result, the eigenvalue problem in Eq.~(\ref{eq:diagonalization problem}) is difficult to solve exactly. Therefore, we use the perturbation expansion with respect to the wave vector $\bm{k}$:
\begin{eqnarray}
\begin{pmatrix}
\tilde{\zeta}^{\rm (i)}_1(\bm{k}) \\
\tilde{\zeta}^{\rm (i)}_2(\bm{k}) \\
\tilde{\zeta}^{\rm (i)}_3(\bm{k}) \\
\tilde{\zeta}^{\rm (i)}_4(\bm{k}) \\
\end{pmatrix}
= 
\begin{pmatrix}
\tilde{\zeta}^{\rm (i),0}_1(\bm{k}) \\
\tilde{\zeta}^{\rm (i),0}_2(\bm{k}) \\
\tilde{\zeta}^{\rm (i),0}_3(\bm{k}) \\
\tilde{\zeta}^{\rm (i),0}_4(\bm{k}) \\
\end{pmatrix}
+ k
\begin{pmatrix}
\tilde{\zeta}^{\rm (i),1}_1(\bm{k}) \\
\tilde{\zeta}^{\rm (i),1}_2(\bm{k}) \\
\tilde{\zeta}^{\rm (i),1}_3(\bm{k}) \\
\tilde{\zeta}^{\rm (i),1}_4(\bm{k}) \\
\end{pmatrix}
+ \cdots,
\end{eqnarray}
\begin{eqnarray}
\lambda_i(\bm{k}) = k \lambda_i^0(\bm{k}) + k^2 \lambda_i^1(\bm{k}) + \cdots,
\end{eqnarray}
and calculate the solution to $O(k^2)$. This approximation is the second one that is used to obtain Eqs.~(\ref{eq:correlation function: integral expression: xx}) and (\ref{eq:correlation function: integral expression: zz}). The calculation of the perturbation expansion is lengthy but straightforward, and so the detailed steps are omitted in this paper.

The solution of Eq.~(\ref{eq:master equation of xi}) is written using the eigenvector as
\begin{eqnarray}
\begin{pmatrix}
\tilde{\xi}_1 \\
\tilde{\xi}_2 \\
\tilde{\xi}_3 \\
\tilde{\xi}_4 \\
\end{pmatrix} = 
\sum_{i=1}^4 a^{\rm (i)}(\bm{k},t)
\begin{pmatrix}
\tilde{\zeta}^{\rm (i)}_1(\bm{k}) \\
\tilde{\zeta}^{\rm (i)}_2(\bm{k}) \\
\tilde{\zeta}^{\rm (i)}_3(\bm{k}) \\
\tilde{\zeta}^{\rm (i)}_4(\bm{k}) \\
\end{pmatrix},
\label{eq:expression of xi intermsof eigenvector}
\end{eqnarray}
and $a^{\rm (i)}(\bm{k},t)$ is given as the solution of the following equation
\begin{eqnarray}
\Biggl(\frac{\partial}{\partial t} - \dot{\gamma} k_x \frac{\partial}{\partial k_z} + \lambda_i(\bm{k}) \Biggr)a^{\rm (i)}(\bm{k},t) = 0.
\label{eq:master equation of a}
\end{eqnarray}
We can solve Eq.~(\ref{eq:master equation of a}) without any approximations to obtain the integral expression for $a^{\rm (i)}(\bm{k},t)$
\begin{eqnarray}
a^{\rm (i)}(\bm{k},t) = a^{\rm (i)}(\bm{k}(-t),0) \exp \Big(-\int_0^t ds  \lambda_i(\bm{k}(-s))\Big) ,\nonumber \\
\label{eq:expression of a intermsof eigenvalue}
\end{eqnarray}
where $\bm{k}(-t)=(k_x,k_y,k_z+\dot{\gamma}tk_x)$. 

For $k_y=k_z=0$, which is discussed in the main text, the vector $\{\hat{\bm{e}}^{\rm (a)}(\bm{k})\}_{a=1,2,3}$ can be simplified as $\hat{\bm{e}}^{\rm (1)}(\bm{k})=(1,0,0)$, $\hat{\bm{e}}^{\rm (2)}(\bm{k})=(0,0,1)$, and $\hat{\bm{e}}^{\rm (3)}(\bm{k})=(0,-1,0)$. Then, we have expressions for $C_{xx}\kx$ and $C_{zz}\kx$ in terms of $\tilde{\bm{\xi}}^{\rm (i)}(\bm{k})$:
\begin{eqnarray}
\hspace{-1cm} \big\langle \tilde{\xi}_2\kx\tilde{\xi}_2(k'_x)\big\rangle &=& C_{xx}\kx \delta(k_x+k'_x),\\
\hspace{-1cm}\big\langle \tilde{\xi}_3\kx\tilde{\xi}_3(k'_x)\big\rangle &=& C_{zz}\kx \delta(k_x+k'_x).
\end{eqnarray}
Thus, Eqs.~(\ref{eq:correlation function: integral expression: xx}) and (\ref{eq:correlation function: integral expression: zz}) are obtained by using Eqs.~(\ref{eq:expression of xi intermsof eigenvector}) and (\ref{eq:expression of a intermsof eigenvalue}) and substituting the explicit forms of the eigenvector and eigenvalue to $O(k^2)$.

\section{Measurement of viscosity}
\label{appendix:Measurement of viscosity}
There are two viscosities, $\eta_0$ and $\zeta_0$, in the fluctuating hydrodynamic equation. We use the Green--Kubo formula to measure them in the MD simulations~\cite{green1954markoff,jaeger2018bulk}. The Green--Kubo formula provides the microscopic expression of the transport coefficient, and is a useful tool for computing the transport coefficient in the MD simulations.

The Green--Kubo formula for the viscosity is given by~\footnote{
Strictly speaking, these Green--Kubo formulae provide the renormalized transport coefficients, which are different from those of the bare transport coefficients, $\eta_0$ and $\zeta_0$, that appear in the nonlinear fluctuating hydrodynamics. However, we neglect this difference because it is known to be quite small~\cite{das2011statistical}.
}:
\begin{eqnarray}
\eta_0 &=& \frac{V}{3k_B T} \sum_{\alpha \beta} \int_0^{\infty} dt \langle \hat{P}_{\alpha \beta}(t)\hat{P}_{\alpha \beta}(0) \rangle , \\
\zeta_0 &=& \frac{V}{k_B T}  \int_0^{\infty} dt \langle \delta\hat{P}(t) \delta\hat{P}(0) \rangle ,
\end{eqnarray}
where $\hat{P}_{\alpha \beta}(t)$ is the microscopic expression of total stress tensor
\begin{eqnarray}
\hspace{-0.5cm} \hat{P}_{\alpha \beta}(t) &=& \frac{1}{V}\Big(\sum_{i=1}^N \frac{p_{i\alpha} p_{i \beta}}{m} + \sum_{i=1}^N \sum_{j>i} (r_{i \alpha} -r_{j \alpha}) f_{ij, \beta}\Big),
\end{eqnarray}
and $\delta \hat{P}(t) = \frac{1}{3} \sum_{\alpha} \hat{P}_{\alpha \alpha}(t) - \big\langle \frac{1}{3} \sum_{\alpha} \hat{P}_{\alpha \alpha}(t)\big\rangle$. The summation in the expression for $\eta_0$ is taken over the off-diagonal elements.

\begin{figure}[t]
\centering
\includegraphics[width=8.6cm]{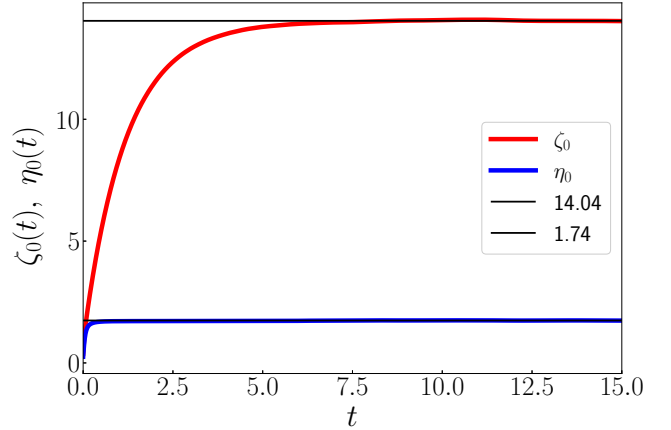}
\caption{Time integral of equilibrium time correlation as a function of $t$. Red: $\zeta_0(t)$. Blue: $\eta_0(t)$. The setup and parameters are the same as in the main text, but the system size is fixed to $L_x=L_y=L_z=64$.}
\label{fig:viscosity.wca}
\end{figure}

Figure~\ref{fig:viscosity.wca} displays the time integral of the equilibrium time correlation function 
\begin{eqnarray}
\eta_0(t) &=& \frac{V}{3k_B T} \sum_{\alpha \beta} \int_0^{t} ds \langle \hat{P}_{\alpha \beta}(s)\hat{P}_{\alpha \beta}(0) \rangle ,\\
\zeta_0(t) &=& \frac{V}{k_B T}  \int_0^{t} ds \langle \delta\hat{P}(s) \delta\hat{P}(0) \rangle 
\end{eqnarray}
for $L_x=L_y=L_z=64$. The setup and parameters of the MD simulation are the same as in the main text. We take an ensemble average over $16$ noise realizations and a time average over $10000$. This figure indicates the existence of a plateau region of the time integral. We adopt the plateau value as the values of $\eta_0$ and $\zeta_0$.

\section{Measurement of relaxation time}
\label{appendix:Measurement of relaxation time}
In the MD simulations, all observations are taken in the nonequilibrium steady state. This state is prepared by a relaxation run lasting about $3$--$10$ times longer than the relaxation time. Here, we explain how to estimate the relaxation time.

\begin{figure*}[th]
\centering
\begin{tabular}{cc}
\begin{minipage}{0.5\hsize}
\begin{center}
\includegraphics[width=8.6cm]{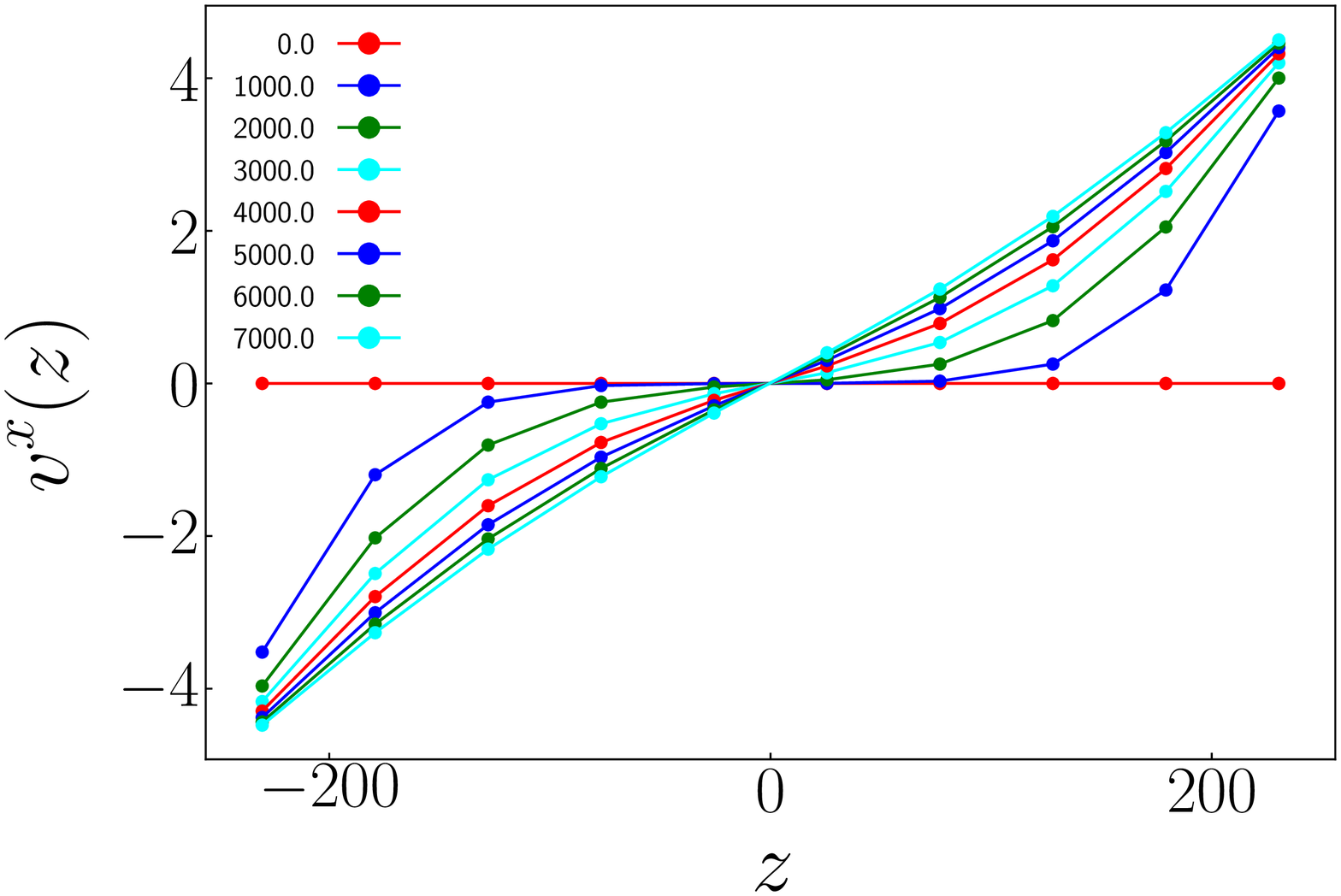} 
\end{center}
\end{minipage}
\begin{minipage}{0.5\hsize}
\begin{center}
\includegraphics[width=8.6cm]{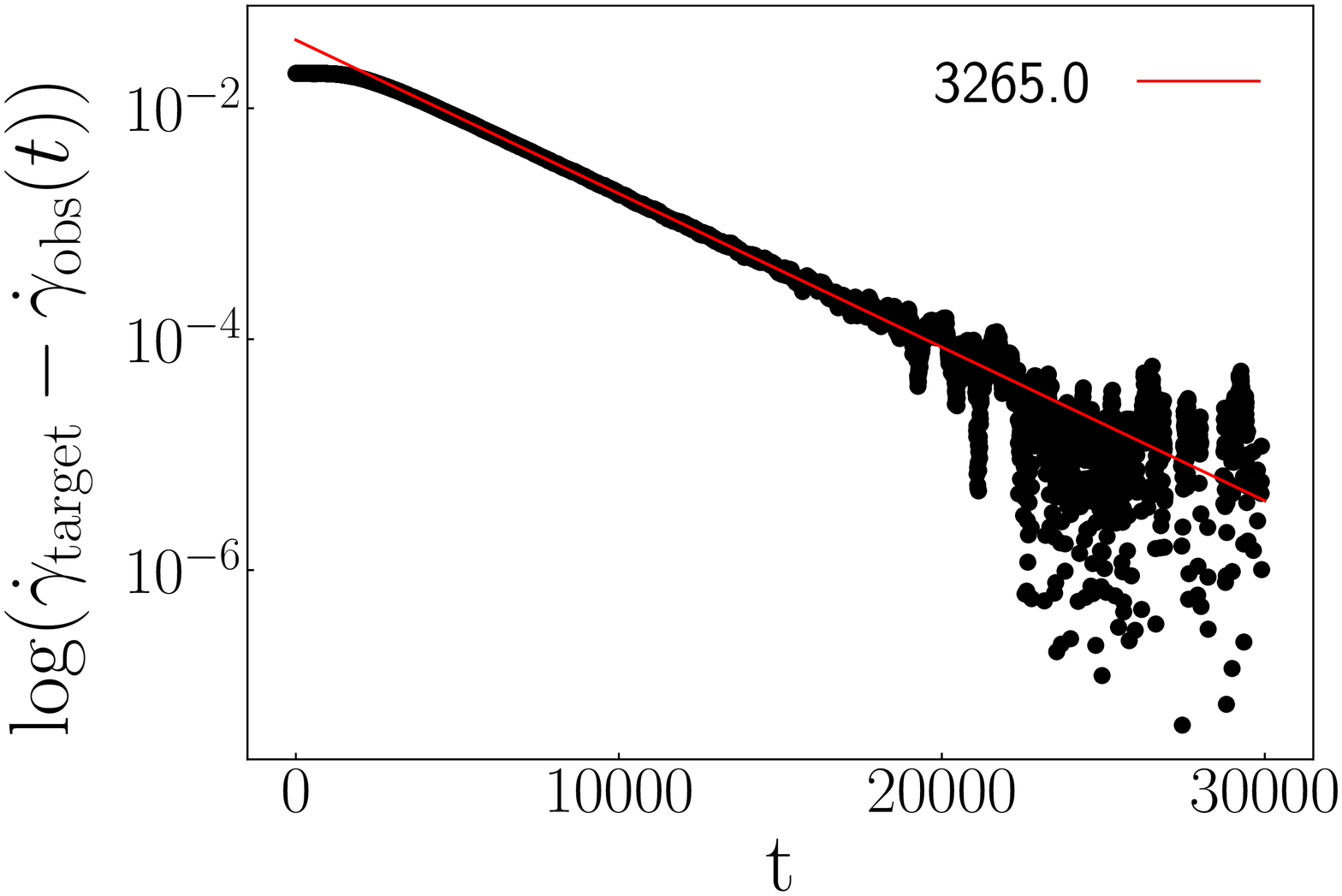} 
\end{center}
\end{minipage}
\end{tabular}
\caption{Relaxation process of velocity field for $L_x=1024$, $L_y=32$, $L_z=512$, and $\dot{\gamma}=0.02$. Left: time evolution of velocity profile $v^x(z)$. Right: $\log(\dot{\gamma}_{\rm obs}(t)-\dot{\gamma}_{\rm target})$ vs $t$}
\label{fig:relaxation process of velocity field}
\end{figure*}

Because the slow variables of our system are the density and momentum, it is reasonable to assume that the relaxation time can be estimated from the relaxation process of the velocity field. Thus, we prepare the initial state in which the particles are randomly located with zero overlaps and their velocities are given according to the uniform distribution with a temperature of $T=1.0$. We run the simulation under the Lees--Edwards boundary condition. The left-hand panel of Fig.~\ref{fig:relaxation process of velocity field} displays the typical relaxation process of the velocity profile $v^x(z)$ for $L_x=1024$, $L_y=32$, $L_z=512$, and $\dot{\gamma}=0.02$. The velocity field relaxes to Eq.~(\ref{eq:velocity profile in the steady state}) after a sufficiently long time.

The gradient of the velocity profile, $\dot{\gamma}_{\rm obs}(t)$, at $z=0$ (the farthest position from the boundaries) is presented in the right-hand panel of Fig.~\ref{fig:relaxation process of velocity field}. From this figure, we find that $\dot{\gamma}_{\rm obs}(t)$ decays to the target shear rate $\dot{\gamma}_{\rm target}=0.02$ in the exponential form
\begin{eqnarray}
\dot{\gamma}_{\rm obs}(t) = \dot{\gamma}_{\rm target} + A_{g}\exp(-t/\tau_{\rm relax}),
\label{eq:exponential decay of shear rate}
\end{eqnarray}
where $\tau_{\rm relax}$ is the relaxation time of the velocity field. The red line in the right-hand panel of Fig.~\ref{fig:relaxation process of velocity field} represents the fitting result using Eq.~(\ref{eq:exponential decay of shear rate}). In this case, the relaxation time $\tau_{\rm relax}$ is estimated as $\tau_{\rm relax}=3265$.

%

\end{document}